\documentclass[journal]{IEEEtran}

\usepackage{ifpdf}
\usepackage{cite}
\usepackage{graphicx}
\usepackage{amssymb}
\usepackage{amsmath,amsfonts}
\usepackage{stmaryrd} 
\usepackage{xcolor}
\usepackage{algorithmic}
\usepackage{textcomp}
\usepackage{bbm}
\usepackage{stmaryrd} 
\usepackage{amsthm}
\usepackage{mathrsfs}
\usepackage{tikz}
\usepackage{setspace}
\usetikzlibrary{automata,arrows,positioning,calc}
\definecolor{celadon}{rgb}{0.67, 0.82, 0.59}
\definecolor{cblue}{rgb}{0.6, 0.73, 0.89}
\newtheorem{thm}{Theorem}
\newtheorem{cor}{Corollary}
\newtheorem{rema}{Remark}
\newtheorem{prop}{Proposition}
\newtheorem{lem}{Lemma}
\newtheorem{defi}{Definition}

\theoremstyle{definition}
\newtheorem{ex}{Example}
\newcommand{\norm}[1]{\left\lVert#1\right\rVert}
\begin{document}
	
	\title{Zero-Error Feedback Capacity \\{for Bounded Stabilization and}\\ Finite-State Additive Noise Channels}
	\author{{Amir~Saberi,~\IEEEmembership{Member,~IEEE,}
		Farhad~Farokhi,~\IEEEmembership{Senior Member,}
		and~Girish~N.~Nair,~\IEEEmembership{Fellow,~IEEE}}
		\thanks{The first author is with the School of Engineering at The Australian National University (e-mail: amir.saberi@anu.edu.au), 2nd and 3rd authors are with the Department of Electrical and Electronic Engineering, The University of Melbourne, VIC 3010, Australia  (e-mails: \{ffarokhi, gnair\}@unimelb.edu.au). This work was partially supported by the Australian Research Council via Future Fellowship grant FT140100527. This paper was presented in part at the 2020 IEEE International Symposium on Information Theory \cite{saberi2020anexplicit}.}}
	
	\markboth{To Appear in IEEE Transactions on Information Theory}%
	{Saberi \MakeLowercase{\textit{et al.}}}
	
	\maketitle
	
	\begin{abstract} 
		This article studies the zero-error feedback capacity of {\em causal} discrete channels with memory. First, by extending the classical zero-error feedback capacity concept, a new notion of {\em uniform zero-error feedback capacity} $ C_{0f} $ for such channels is introduced. Using this notion a tight condition for bounded stabilization of unstable noisy linear systems via causal channels is obtained, assuming no channel state information at either end of the channel.
		
		Furthermore, the zero-error feedback capacity of a class of additive noise channels is investigated. It is known that for a discrete channel with correlated additive noise, the ordinary capacity with or without feedback is equal $ \log q-\mathcal{H}_{ch} $, where $ \mathcal{H}_{ch} $ is the entropy rate of the noise process and $ q $ is the input alphabet size. In this paper, for a class of finite-state additive noise channels (FSANCs), it is shown that the zero-error feedback capacity is either zero or $C_{0f} =\log q -h_{ch} $, where $ h_{ch} $ is the {\em topological entropy} of the noise process. A condition is given to determine when the zero-error capacity with or without feedback is zero. This, in conjunction with the stabilization result, leads to a ``Small-Entropy Theorem'', stating that stabilization over FSANCs can be achieved if the sum of the topological entropies of the linear system and the channel is smaller than $\log q$.
	\end{abstract}
	
	\section{Introduction}
	
	The {\em zero-error capacity} $ C_0 $ is defined as the maximum block coding rate yielding exactly zero decoding errors at the receiver. Although this concept was introduced by Shannon over 60 years ago \cite{shannon1956zero}, the general formula for it is still missing, even for memoryless channels (see \cite{korner1998zero} for a detailed survey). However, Shannon derived a single-letter formula for the {\em zero-error feedback capacity} $ C_{0f} $ of a discrete memoryless channel (DMC) with noiseless feedback. In recent years, there has been some progress towards determining $ C_{0f} $ for channels with an internal state. In \cite{zhao2010zero}, a dynamic programming formulation for computing $ C_{0f} $ was introduced for a {\em Finite-State Channel}~(FSC) modelled as a Markov decision process assuming that state information is available at both encoder and decoder. In \cite{bracher2017zero}, Gel’fand-Pinsker channels having i.i.d. internal states are studied, and single-letter formula for $ C_{0f} $ are derived assuming that the states are known at the transmitter. However, the problem is still open when there is memory in the state process with no state information at the encoder or decoder. This may be the case, for instance, if the channel is being used for controlling an unstable system in real-time, where there may not be enough time to obtain reliable estimates of the channel state in between channel uses.  
	
	Indeed, the impact of channel capacity on state estimation and control has been a major topic of research in the control and information theory literature for the last two decades. One of the fundamental problems in this field is to find conditions such that remote state estimation or control is feasible over a given channel, e.g., \cite{elia2004bode, matveev2009estimation,Franceschetti2014}. In other words, for the estimation problem, the receiver should reconstruct the current state of the remote system in real-time, and for the control problem, it should calculate a control input sequence to stabilize the system. Major studies have considered control problems over noiseless communication channels, e.g., \cite{brockett2000quantized, tatikonda2004controla, nair2004stabilizability}. In this case, the ``data rate theorem'' states that state estimation or control over a memoryless channel is feasible if and only if (iff) the average channel bit-rate is larger than the {\em topological entropy} of the linear system. This holds with or without plant noises and under different notions of stability or convergence,
	such as $r - $th moment, uniformly or almost-surely \cite{ tatikonda2004controla, nair2004stabilizability, wong1997systems,wong1999systems,nair2003exponential}. The topological entropy measures the asymptotic growth rate of uncertainty in a dynamical system and was first introduced by Adler \emph{et. al.} \cite{adler1965topological}. For linear systems with dynamical matrix $ A $, it reduces to the sum of the logarithms of the unstable eigenvalues of $ A $. In recent years, further connections between the topological entropy of dynamical systems and information theory have been unveiled, e.g., \cite{kawan2018optimal,liberzon2017entropy}. Recent work has also considered plants with randomly varying dynamical parameters in addition to additive noise \cite{kostina2021stabilizing}.
	
	Extensions of the data rate theorem to noisy communication channels are considered for discrete channels in \cite{tatikonda2004controlb, sahai2006necessity, matveev2009estimation, matveev2007analogue, nair2013nonstochastic} and for Gaussian channels in \cite{elia2004bode,braslavsky2007feedback}. Depending on the stability notion being considered, the data-rate in the theorem is replaced with either the ordinary capacity \cite{tatikonda2004controlb, matveev2007analogue}, the {\em anytime capacity}  \cite{sahai2006necessity}, or the zero-error capacities with and without feedback \cite{matveev2007shannon, matveev2009estimation, nair2013nonstochastic}. Tight results on control performance have also been obtained in a mean-square setting \cite{8693967}.
	
	When estimating states of a linear system over a noisy DMC, for achieving estimation errors that are almost-surely {\em uniformly bounded}, it is known that $ C_0 $ of the channel has to be larger than the topological entropy of the linear system \cite{matveev2007shannon, matveev2009estimation}. Furthermore, for achieving bounded states that are almost-surely uniformly bounded, the border for stability is the zero-error feedback capacity $ C_{0f} $ of such channels \cite{matveev2007shannon, matveev2009estimation, Franceschetti2014}. In \cite{nair2013nonstochastic}, a nonstochastic framework is proposed and used to show that for the estimation problem over memoryless channels, $ C_0 $ is still the figure of merit for achieving uniformly bounded estimation errors, having no stochastic assumptions on the uncertainties. 
	
	Despite the extensive literature above, few studies have considered these problems over channels with memory. The notable exceptions are the study of mean-square stabilization over a moving average Gaussian channel in \cite{middleton2009feedback} and over an autoregressive Gaussian channel in \cite{charalambous2020ergodic}. {In \cite{saberi2020bounded}, we showed that the boundary for bounded state estimation over FSCs is  $ C_0 $. However, the stabilization problem was remained open.} This paper intends to fill this gap by revealing tight conditions for uniformly bounded stabilization over channels with memory. 
	
	Shannon showed that the stochastic structure of a DMC is not needed to obtain its zero-error capacity (with or without feedback), and it is enough to know the possible set of outputs for each channel input letter. This suits settings where the channel noise distributions are not available or where the noise is not random. In such situations, worst-case analysis is more relevant \cite{wang2018end, badr2017layered, lim2017information, wang2019large,wiese2018secure}, which is equivalent to the presence of an {\em omniscient} adversary who knows the transmitted codeword a priori and can input malicious noise to the channel \cite{wang2019large}. Here, we use the {\em uncertain variable} framework introduced in \cite{nair2013nonstochastic} for studying the zero-error capacity. This allows us to work with ranges of the variables in the problem rather than their probability distributions. This approach has shown to be useful in several setups, e.g., \cite{Rangi2019non,gagrani2020worst}. However, the results readily apply to the channels with known stochastic information.
	
	\subsection{Our contribution}
	In the first part of this paper, {\em causal} channels and their zero-error capacities are considered. The concept of a {\em uniform zero-error feedback code} is introduced, extending the classic definition of a {\em zero-error feedback code}. We show that for an FSC both definitions are equivalent (Proposition \ref{prop:fsc0f}). These concepts are utilized in the subsequent sections.
	
	We then turn our focus on studying the uniformly bounded {stabilization} of a linear time-invariant (LTI) system over causal channels (introduced in the first part). We show that to achieve stability, it is necessary that $ C_{0f} $ of the channel is equal or larger than the topological entropy $ h_{lin} $ of the LTI system (Theorem \ref{thm:stab}). If this condition does not hold, then there is no encoder-controller pair that can stabilize the system.  This result holds even if there is no explicit feedback from the channel output back to the encoder, apart from through the linear system. This is a significant generalization of the stabilization result of \cite{matveev2007shannon}, which is restricted to discrete memoryless channels. {We also show that this result is tight, i.e., if $ C_{0f} > h_{lin} $ there is a coder and controller that achieves bounded stabilization. In other words, the bounded stabilization condition is of the form
		\begin{align*}
			C_{0f} \gtrsim h_{lin}.
		\end{align*}
		Here, we adopt the notation used in \cite{Franceschetti2014} where the symbol $ \gtrsim $ is used to indicate that the inequality is strict for the sufficient but not for the necessary condition. Therefore, we not only settle the conjecture in \cite{nair2012nonstochastic}, which was stated for memoryless channels, but also we extend it to channels with memory.} The achievability argument adapts the approach of \cite{matveev2009estimation}. However, we do not need to use their random coding technique. This result is also analogous to our previous work on linear state estimation over channels with memory  \cite{saberi2019state,saberi2020bounded}, where we showed that bounded estimation errors could be achieved if $C_0 > h_{lin}$. However, the feedback interconnection here between the channel and the LTI system requires different proof techniques.
	
	We next return to the zero-error capacity problem and consider $ q- $ary channels with additive noise generated by a {\em finite-state machine}. We prove that when $ C_0 \neq 0 $,
	\begin{align*}
		C_{0f} = \log q - h_{ch},
	\end{align*}
	where  $ h_{ch} $ is the topological entropy of the channel noise (Theorem \ref{thm:zero-error}).
	Topological entropy for discrete systems is borrowed from the symbolic dynamics literature and is defined as the asymptotic growth rate of the number of possible state sequences in a finite-state machine \cite{lind1995introduction}. This result is a zero-error analogue of the  formula $ C_f = \log q - \mathcal{H}_{ch}$ \cite{alajaji1995feedback} for the ordinary feedback capacity of stochastic additive noise channels with Shannon entropy rate $\mathcal{H}_{ch}$. 
	In previous work on state estimation, we have shown that $ \log q - h_{ch} $ upper bounds $ C_{0f} $ \cite{saberi2020bounded}. Here we show that this upper bound is, in fact, achievable.\footnote{For the sake of completeness, the proof of the $ C_{0f} $ converse is given in Appendix \ref{app:zero-error}.} Unlike  \cite{bracher2017zero,zhao2010zero}, we do not assume that channel state information is available at the encoder or decoder. Examples, including a Gilbert-Elliott channel, are considered for which the explicit value of $ C_{0f} $ is computed. 
	
	Finally, combining the previous results, we show that a tight condition for uniformly bounded stabilization via such channels is that
	\begin{align*}
		h_{lin} + h_{ch} \lesssim \log q.
	\end{align*}
	In other words, closed-loop stabilization can be achieved if the sum of the topological entropies of the two interconnected systems is smaller than the number of bits transmittable per channel use. Moreover, if this condition does not hold, then no coding and control scheme can stabilize the system. Here, both continuous and discrete topological entropies appear in one equation. 
	
	The results presented in this paper go well beyond our preliminary work reported in \cite{saberi2020anexplicit} as well as our work on state estimation in \cite{saberi2020bounded, saberi2019state}.  This paper includes new results on general causal channels and stabilization in addition to the proof of Theorem \ref{thm:c0f0} and new properties discussed in Section \ref{sec:apc0}.
	
	\subsection{Paper organization}
	The rest of the paper is organized as follows. In Section \ref{sec:ZN}, causal channels are introduced in a nonstochastic framework, and their properties are investigated. In Section \ref{sec:stab}, necessary and sufficient conditions for having uniformly bounded estimation and stabilization over general communication channels with memory is obtained. In Section~\ref{sec:model}, the FSANC model and associated zero-error capacity results are presented. Moreover, some examples of FSANCs, including a Gilbert-Elliott channel, are discussed in this section. The results of the previous sections are combined in Section \ref{sec:stabofsan} and a tight condition for stabilization over FSANCs is presented.
	Finally, concluding remarks and future extensions are discussed in section~\ref{sec:conclusions}.
	
	\subsection{Notation}
	Throughout the paper, calligraphic letters, such as $ \mathcal{X} $, denote sets. The cardinality of set $ \mathcal{X} $ is denoted by $ \big|\mathcal{X}\big| $. With a slight abuse of notation, we use the same notation to also denote absolute value of a scalar variable. The channel input alphabet size is $ q $, logarithms are in base $2$. Symbols $\oplus$ and $ \ominus $ are modulo $ q $ addition and subtraction respectively. Random (or uncertain) variables are denoted by upper case letters, such as $ X $, and their realizations are denoted by lower case letters, such as $ x $. $ \mathbf{B}_l $ is an $ l $-ball $ \{ v: \norm{v} \leq l\} $ centred at origin with $ \norm{\cdot} $  denoting a norm on a finite-dimensional real vector space. The sequence {$ \{x_i\}_{i=m}^n $} is denoted by $ x_{m:n} $ and $ x_{t_0+[m:n]} $ denotes {$ \{x_i\}_{i=t_0+m}^{t_0+n} $}. Further, if there is no ambiguity in time segment, it is denoted by a vector $ \mathbf{x} $. Also, the semi-infinite sequence where $ n=\infty $ is denoted by $ x_{m:\infty} $.

	\section{Causal Channels and Zero-error Capacities} \label{sec:ZN}
	In this section, we use the {\em uncertain variable} framework introduced in \cite{nair2013nonstochastic} for studying the zero-error capacity. This allows us to work with ranges of the variables rather than their probability distributions. However, the results also apply to the channels with a known probabilistic model. First, the necessary definitions and tools are given, and then the concept of causality in channels and the zero-error capacity with and without feedback are defined. Next, the zero-error feedback capacity of the causal channel is revisited and uniform zero-error feedback codes are introduced, facilitating the connection of the zero-error communication with feedback to applications that require repetitive use of a code such as control systems. Such codes extend the classical zero-error feedback code concept to general causal channels. It is shown that these concepts are equivalent when applied to {FSC}s.
	
	\subsection{Definitions and formulations} \label{sec:deff}
	Let $\Pi$ be a sample space. An uncertain variable $X$ is a mapping from $ \Pi $ to a set  $\mathcal{X}$. 
	Given other uncertain variable $Y$,  the {\em marginal}, {\em joint} and {\em conditional ranges} are denoted 
	\begin{align*}
		\llbracket X\rrbracket:=& \{ X(\pi) : \pi \in \Pi \} \subseteq \mathcal{X}, \\
		\llbracket X,Y\rrbracket:=& \{ (X(\pi),Y(\pi)) : \pi \in \Pi \} \subseteq \llbracket X \rrbracket\times \llbracket Y \rrbracket,\\
		\llbracket Y|x\rrbracket:=& \{ Y(\pi) : X(\pi)=x, \pi \in \Pi \} \subseteq \llbracket Y \rrbracket. 
	\end{align*}
	Consider uncertain variables $X$, $ Y $ and $Z$.  We say $ X $ and $ Y $ are {\em mutually unrelated} if $\llbracket X,Y \rrbracket = \llbracket X \rrbracket \times \llbracket Y \rrbracket$, i.e., if the joint range is the Cartesian product of the marginal ranges. In the context of random variables, this is analogous to the {\em support} of the joint distribution of $ X $ and $ Y $ equaling the Cartesian product of the marginal supports, without saying anything about whether the joint distribution itself factorizes. This is equivalent to the conditional range property $\llbracket X|y \rrbracket = \llbracket X \rrbracket$, $\forall y\in \llbracket Y \rrbracket$ (cf. \cite[Lem. 2.1]{nair2013nonstochastic}).
	Moreover, $ X $,  $ Y $, and $ Z $ are said to form a {\em Markov uncertainty chain} if 
	\begin{align*}
		\llbracket X|y,z \rrbracket = \llbracket X|y \rrbracket, \forall (y,z)\in \llbracket Y,Z \rrbracket.
	\end{align*}
	\begin{defi}[Zero-error code] \label{zecode} 
		A set $ \mathcal{F} {\subseteq} \mathcal{X}^n $ is a $ (|\mathcal{F}|,n) $ zero-error code for a channel if no two distinct codewords $ x_{1:n} \neq x'_{1:n} \in \mathcal{F}$  can  result in the same channel output sequence, i.e., $  \llbracket Y_{1:n}|x_{1:n} \rrbracket \cap  \llbracket Y_{1:n}|x'_{1:n} \rrbracket  =\emptyset$.
	\end{defi}
	
	\begin{rema} \label{rem:dist}
		Consider the conditional range $\llbracket X_{1:n}|y_{1:n} \rrbracket$, which is the set of inputs that can produce output $ y_{1:n} $. This is equivalent to the {\em adjacent} inputs for a given output for a discrete channel.\footnote{Two channel inputs $ x_1 $ and $ x_2 $ are adjacent if they can produce {the} same output $ y $ \cite{shannon1956zero}.} Moreover, if two inputs are non-adjacent, i.e., $ x_{1:n} \in \llbracket X_{1:n}|y_{1:n} \rrbracket $  and $ x'_{1:n} \in \llbracket X_{1:n}|y'_{1:n} \rrbracket $ such that $  \llbracket X_{1:n}|y_{1:n} \rrbracket \cap  \llbracket X_{1:n}|y'_{1:n} \rrbracket =\emptyset $ then {these} two inputs can be distinguished unambiguously at the decoder.
	\end{rema}
	The zero-error capacity of a channel is defined as follows.
	\begin{defi} [Zero-error capacity] \label{def:c0}
		The zero-error capacity of a channel is
		\begin{align}
			C_0:=\sup_{n \in \mathbb{N}, \, \mathcal{F} \in \mathscr{D}(n)}
			\frac{\log |\mathcal{F}|}{n}, \label{c0def}
		\end{align}
		where $ \mathscr{D}(n)\subseteq \mathcal{X}^{n}$  is the set of all zero-error codes of length $ n $.
	\end{defi}
	{\begin{rema}
		In a probabilistic setting, zero-error codes and capacity are more stringent than the standard ordinary or ``small-error'' notions. The latter allow for an arbitrarily small probability of decoding error, whereas the former require that the probability of decoding error be exactly zero.
	\end{rema}}
	The { \em zero-error feedback code} is defined in the presence of noiseless feedback from the channel output. Hereafter, we require channels to be {\em causal}, i.e., the current output does not depend on the future values of the input \cite{massey1990causality}.
	\begin{defi}[Causal channel]\label{def:causc}
		A causal channel is a mapping from input sequences $ X_{1:\infty}$ and noise sequence $ Z_{1:\infty}$ to an output sequence $ Y_{1:\infty}$, where $X_i \in \mathcal{X} $, $Z_i \in \mathcal{Z} $ and $ Y_i \in \mathcal{Y} $ such that the output at each time depends only on the past and current inputs as well as a noise sequence, i.e.,
		\begin{align}
			Y_i &= g_i(X_{1:i},Z_{1:i}),\, i \in \mathbb{N}, \label{causc}
		\end{align} 
		where $ Z_i $ is an uncertain variable, mutually unrelated to $ (X_{1:i},Z_{1:i-1}) $.
	\end{defi}
	
	We define a {\em zero-error feedback code} in the following by assuming that the input is a function of the message and past channel outputs.
	\begin{defi}[Uniform zero-error feedback code] \label{def:zfc} 
		For a causal channel with feedback, the input at each time is a function of the message and previous outputs, i.e., 
		\begin{align}
			X_{t_0+i}=f_{i}(m, Y_{t_0+[1:i-1]}), \, i \in \{1,\dots, n\}, m \in \mathcal{M}, \label{zef}
		\end{align}
		where the transmission starts at time $t_0+1 \in \mathbb{N}$, and $ n \in \mathbb{N} $ is the blocklength. Moreover, $ f_{i}: \mathcal{M}\times\mathcal{Y}^{i-1} \to \mathcal{X}$ is the encoding function. The set of functions $ \mathcal{F}=\{f_{1:n}(m,\cdot)|m\in \mathcal{M}\} $ is called a $ (|\mathcal{M}|,n) $ {\em uniform zero-error feedback code} if for all starting times, no two distinct messages $ m \neq m'  \in \mathcal{M} $ result in the same channel output sequence, i.e.,
		\begin{align*}
			\llbracket Y_{t_0+[1:n]}|m \rrbracket \cap  \llbracket Y_{t_0+[1:n]}|m' \rrbracket  =\emptyset,\, \forall t_0 \in \mathbb{N}\cup\{0\}.
		\end{align*}	
		If this holds only for $ t_0=0 $, {i.e., $\llbracket Y_{1:n}|m \rrbracket \cap  \llbracket Y_{1:n}|m' \rrbracket  =\emptyset$}, then $ \mathcal{F} $ is a {\em zero-error feedback code}.
	\end{defi}
	The uniform zero-error feedback code is more restrictive than the well-known {\em zero-error feedback code} used in literature where the transmission always starts at time 1, i.e., $ t_0=0 $; e.g., \cite{zhao2010zero,bracher2017zero,ahlswede1973channels}. However, this only implies that the encoding strategy does not change over time, and this condition is imposed to be able to use the zero-error feedback code starting from any time $ t_0 +1 \in \mathbb{N}$ or in repetition. This gives the code a {\em uniform} (or {\em time-invariance}) property, and yet the code must be able to convey the messages without error for the corresponding block. For the case where the channel is memoryless, any zero-error feedback code is uniform as the transmissions are unrelated. We also show that for FSCs, any zero-error feedback code is also uniform (Proposition \ref{prop:fsc0f}).
	\begin{defi}[Zero-error feedback capacity]\label{def:c0f}
		The zero-error feedback capacity of a causal channel is 
		\begin{align*}
			C_{0f}:=\sup_{n \in \mathbb{N}, \,  \mathcal{F} \in  \mathscr{F}(n)}	\frac{\log |\mathcal{M}|}{n},
		\end{align*}
		where $\mathcal{F}$ is chosen from the set $ \mathscr{F}(n)$ of all uniform zero-error feedback codes of size $ n $.
	\end{defi}
	\begin{lem}\label{lem:cc0f}
		For a causal channel,
		\begin{align}
			C_{0f}=\lim_{n \to \infty} \sup_{\mathcal{F} \in  \mathscr{F}(n)}
			\frac{\log |\mathcal{M}|}{n}. \label{limc0f}
		\end{align}
	\end{lem}
	\begin{proof}
		Define $ \mathbf{con}(\mathcal{F}_1,\mathcal{F}_2) $ as the concatenation of two uniform zero-error feedback codes $ \mathcal{F}_1 \in  \mathscr{F}(n)$ and $ \mathcal{F}_2 \in  \mathscr{F}(r)$ with message sets $ \mathcal{M}_1 $ and $ \mathcal{M}_2 $, respectively. {In other words, $ \mathcal{F}_1 $ and $ \mathcal{F}_2 $ are applied in succession forming a communication of $ n+r $ channel uses. } The new coding, $ \mathbf{con}(\mathcal{F}_1,\mathcal{F}_2) $ is a $ (|\mathcal{M}_1| \times |\mathcal{M}_2|, n+r) $ uniform zero-error feedback code since it yields distinguishable messages by the uniform property of each code. {Note that by Definition \ref{def:causc}, the last $ r $ transmissions in the channel {do} not affect the first $ n $ transmissions.}	
		Let $ a_k:= \sup_{\mathcal{F} \in  \mathscr{F}(k)}\log |\mathcal{M}|  $,  $ k \in \mathbb{N} $. We have	
		\begin{align*}
			a_{n+r} &= \sup_{\mathcal{F} \in  \mathscr{F}(n+r)}\log |\mathcal{M}|\\
			&\stackrel{(a)}{\geq} \sup_{\mathbf{con}(\mathcal{F}_1,\mathcal{F}_2) \in  \mathscr{F}(n+r)} \log \big(|\mathcal{M}_1|\times |\mathcal{M}_2|\big)\\
			&= \sup_{\mathcal{F}_1\in \mathscr{F}(n),\mathcal{F}_2 \in  \mathscr{F}(r)} \log \big(|\mathcal{M}_1|\times |\mathcal{M}_2|\big)\\
			&=\sup_{\mathcal{F}_1\in \mathscr{F}(n)} \log |\mathcal{M}_1|+\sup_{\mathcal{F}_2\in \mathscr{F}(r)} \log |\mathcal{M}_2|\\
			&=a_n+a_r.
		\end{align*}
		Here, $ (a) $ follows because we have restricted the uniform zero-error feedback code to a special structure of $ \mathbf{con}(\mathcal{F}_1,\mathcal{F}_2) $. This means that the channel is {\em superadditive} and because \mbox{$ a_k/k \leq \log |\mathcal{X}| $}, by Fekete's lemma \cite[Ch. 2.6]{schrijver2003combinatorial}, 
		\begin{align*}
			\sup_{k \in \mathbb{N}}	\frac{\log a_k}{k} &=\lim_{k \to \infty} \frac{\log a_k}{k}.
		\end{align*}
		Thus, it yields \eqref{limc0f}.
	\end{proof}
	\begin{lem} \label{lem:delc0f}
		For a causal channel with $ C_{0f}>0$, and any \mbox{$ 0 < \delta < C_{0f}$ }, there exists a uniform zero-error feedback code with rate \mbox{$ R= C_{0f}-\delta $} for a sufficiently large blocklength.
	\end{lem}
	\begin{proof}
		By assumption, there is at least a uniform zero-error feedback code with a positive rate $R_0\leq C_{0f}$ and blocklength $n$. It is reasonably easy to build a code with a smaller rate than $R_0$. Furthermore, the proof for arbitrarily small $\delta$ follows from Lemma \ref{lem:cc0f}.
	\end{proof}
	\begin{defi}[Finite-state channels]
		\label{def:fschannel}	
		A discrete (time-invariant) finite-state channel has a finite set of states $ \mathcal{S} $ such that the current output $ Y_i $ and next state $ S_{i+1} $ only depend on current input and state, i.e., $ \forall i \in \mathbb{N} $,
		\begin{align}
			S_{i+1}  &= \mathfrak{f} (S_{i},X_{i},\varPhi_i), \label{fssc}\\
			Y_i  &= \mathfrak{h}(S_{i},X_{i},\varPsi_i), \label{fsyc}
		\end{align}
		where {$ \varPhi_i \in \llbracket \varPhi_i \rrbracket = \mathcal{N}_s $,  $ \varPsi_i \in \llbracket \varPsi_i \rrbracket = \mathcal{N}_o$ are the process and measurement noises of the channel, respectively. It is assumed that $ (\varPhi_i,\varPsi_i) $ is mutually unrelated to $ (\varPsi_{1:i-1},\varPhi_{1:i-1},X_{1:i},S_1) $, where $ S_1 $ is the initial state of the channel.} Here, $ \mathfrak{f}:\mathcal{S} \times \mathcal{X} \times \mathcal{N}_s \to \mathcal{S}  $ and $ \mathfrak{h}:\mathcal{S} \times \mathcal{X} \times \mathcal{N}_o \to \mathcal{Y} $ are the state update and the output mappings, respectively.
	\end{defi}
	Further, assuming that the message $ m $ is unrelated to channel noises and initial state, i.e., $ (\varPsi_{1:i},\varPhi_{1:i},S_1) $, relationships \eqref{fssc}-\eqref{fsyc} imply that
	\begin{align}
		\llbracket S_{i+1} | x_{1:i},s_{1:i}, y_{1:i-1}, m \rrbracket &= \llbracket S_{i+1}| x_i,s_i \rrbracket = \llbracket\mathfrak{f} (s_{i},x_{i},\varPhi_i) \rrbracket, \label{stateun}\\
		\llbracket Y_i| x_{1:i},s_{1:i},y_{1:i-1}, m \rrbracket &= \llbracket Y_i| x_i,s_i \rrbracket= \llbracket\mathfrak{h} (s_{i},x_{i},\varPsi_i) \rrbracket,\label{outun}
	\end{align} 
	for admissible sequences of $ x_{1:i},s_{1:i},y_{1:i-1},m $.
	\begin{lem}\label{lem:canc}
		{FSC}s are causal.
	\end{lem}
	\begin{proof} See Appendix \ref{app:fsccaus}.
	\end{proof}
		\tikzstyle{block1} = [draw, fill=white, rectangle, 
	minimum height=3em, minimum width=5em]
	\tikzstyle{block2} = [draw, fill=white, rectangle, 
	minimum height=3em, minimum width=5em]
	\tikzstyle{input} = [coordinate]
	\tikzstyle{output} = [coordinate]
	\tikzstyle{pinstyle} = [pin edge={to-,thick,black}]
	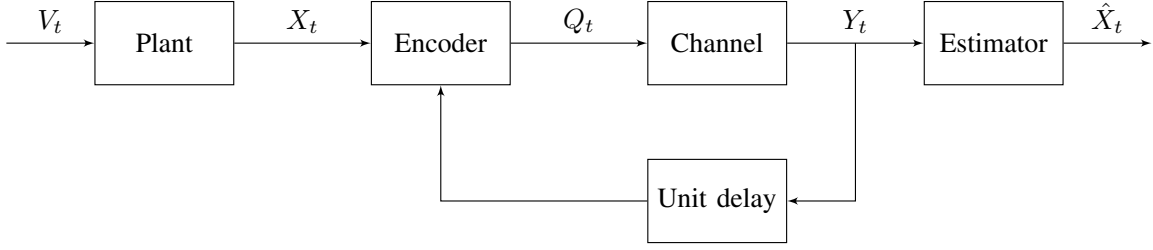
\begin{figure*}[t]
		\centering
		\scalebox{1.05}{\begin{tikzpicture}[auto, node distance=2cm,>=latex']
				\node [input, name=input] {};
				\node [block2, right of=input, node distance=2cm] (plant) { Plant };
				\node [block1, right of=plant, node distance=3.5cm] (encoder) { Encoder };
				\node [block2, right of=encoder, node distance=3.5cm] (channel) { Channel };
				\node [block1, right of=channel, node distance=3.5cm] (decoder) { Estimator} ;
				\node [block1, below of=channel, node distance=2cm] (delay) { Unit delay} ;
				\node [output, right of =decoder, node distance=2cm] (output){};
				
				\draw [draw,->] (input) -- node [name=w]{$ V_t $} (plant);
				\draw [draw,->] (plant) -- node {$ X_t $} (encoder);
				\draw [->] (encoder) -- node {$ Q_t $} (channel);
				\draw [->] (channel) -- node [name=y] {$ Y_t $} (decoder);
				\draw [->] (y) |- (delay);
				\draw [->] (delay) -| (encoder);
				\draw [->] (decoder) -- node  {$ \hat{X}_t $}(output);
		\end{tikzpicture}}
		\caption{\small{State estimation via a communication channel with feedback.}}
		\label{fig:estf}
	\end{figure*}
	In Definition \ref{def:c0f}, the uniform zero-error feedback code has this condition that the messages must be distinguishable for any starting time $ t_0 +1 \in \mathbb{N} $. In the following proposition, we show that any zero-error feedback code for FSCs is also uniform. 
	\begin{prop}\label{prop:fsc0f}
		For FSCs with {an} unknown initial state, any zero-error feedback code is also a uniform zero-error feedback code.
	\end{prop}
	\begin{proof}
		See Appendix \ref{app:fsc0f}.
	\end{proof}	
	\section{Bounded Stabilization over communication channels} \label{sec:stab}
	In this section, we consider the control problem over a causal channel (Definition \ref{def:causc}).
	In communications and networking, it is well understood that most practical channels exhibit memory. However, except for some limited works, this has been often ignored in control systems; see \cite{matveev2009estimation, Franceschetti2014} and references therein. In \cite{tatikonda2004controlb}, it was shown that the Shannon capacity has to be larger than the topological entropy of the plant for almost surely asymptotically stabilization. This condition was shown to be loose for at least DMCs in \cite{matveev2005comments}, which led to establishing that the zero-error capacity of the DMC is the necessary and sufficient border of almost surely asymptotically stabilization in \cite{matveev2007shannon}.  Inspired by this work, in \cite{nair2012nonstochastic}, it is conjectured that for keeping the states of a linear system uniformly bounded over a DMC, the zero-error feedback capacity of the channel has to be larger than the topological entropy of the linear system. Here, we prove this claim not only for DMCs but also for channels with memory. We emphasize that our focus here is on discrete channels. However, the stabilization problem over some special cases of continuous alphabet channels have also studied, e.g., a moving average Gaussian channel is studied in \cite{middleton2009feedback}.
	
	In the sequel, first some preliminaries are given for control systems in Section \ref{sub:prel}. In Section \ref{sub:est}, a tight condition for state estimation of a linear system with uniformly bounded estimation errors over causal channels (Definition \ref{def:causc}) with unit-delay feedback is derived. The structure of this problem is shown in Fig. \ref{fig:estf}. The plant is a dynamic system and is affected by process noise.
	The measured data are quantized and transmitted via the communication channel to the estimator that wishes to reconstruct the states of the plant in real-time. 
	
	This result is utilized to derive conditions for uniformly bounded stabilization over such channels (without feedback) {in Section \ref{sub:ctrl}}. Figure \ref{fig:ctrl} shows the structure of this problem where the controller generates control inputs at each time based on the received channel outputs to stabilize the plant. 
	\subsection{Dynamical system and quantization} \label{sub:prel}
	First, we give the following definition.
	\begin{defi}[Uniform boundedness]
		{A sequence of uncertain variables $ \{X_t (\pi) \in \mathbb{R}^{m}| \pi \in \Pi \}_{t \in \mathbb{N}} $ is {\em uniformly bounded} if $ \exists l >0 $ such that for any initial condition range $ \llbracket X_1 \rrbracket \subseteq \mathbf{B}_l $ and  realization $ x_1 \in \llbracket X_1 \rrbracket $, 
			\begin{align*}
				\sup_{t\in \mathbb{N}, \pi \in \Pi}\norm{X_t(\pi)} =\sup_{t \in \mathbb{N} } \sup \llbracket  \|X_t\| \rrbracket <\infty
			\end{align*}
			with respect to some norm $ \norm{\cdot} $ in $ \mathbb{R}^{m} $, where the inner supremum is also over all valid noise realizations.  The radius $ l $ of the ball $ \mathbf{B}_l $ are known at both ends of the communication channel.}
	\end{defi}
	Consider a linear time-invariant (LTI) dynamical system 
	\begin{align}
		X_{t+1}&=AX_t+BU_t+V_t \in\mathbb{R}^{n_x}, \label{lti1}
	\end{align}
	where $A$ and $B$ are constant matrices, $X_t \in \mathbb{R}^{n_x},\, U_t \in \mathbb{R}^{n_u}$, and the uncertain variable $ V_t \in \mathbb{R}^{n_x}$ represent the process states, control input, and {noise}, respectively. Here, the goal is to keep the estimation error $ (\hat{X}_t-X_t) $, uniformly bounded with $ \hat{X}_t$ denoting the state estimate based on the measurement sequence $Y_{1:t-1}$. The following assumptions are made:
	\begin{itemize}
		\item[A1:] There exist uniform bounds on the initial condition and {the noise at all times}, i.e.,  $\norm{X_1} \leq D_x$ and $\norm{V_t} \leq D$, {$ \forall t \in \mathbb{N} $};
		\item[A2:] The initial state $X_1$, the noise signal $V_t, \, t\in \mathbb{N}$, and the channel error patterns are {\em mutually unrelated};
		\item[A3:]  The zero-noise sequence is valid{, i.e., $V_t=0, \forall t \in \mathbb{N}$, is a possible noise sequence};
		\item[A4:] $A$ has one or more eigenvalues $\lambda_i$ with magnitude greater than one;
		\item[A5:] {The pair $ (A,B) $ is {\em stabilizable}, i.e., the unstable states of the LTI system are {\em controllable}.}
	\end{itemize}
	
	The {\em topological entropy} of the system is given by 
	\begin{align*}
		h_{lin} := \sum_{|\lambda_i|\geq 1} \log|\lambda_i|,
	\end{align*}
	and can be viewed as the rate at which it generates uncertainty. 
	
	\begin{defi}[Contraction quantizer]\label{def:rcont}
		An $ M- $level quantizer $ \mathscr{Q} $ in $ \mathbb{R}^{n_x} $ is a partition of the unit ball $ \mathbf{B}_1 $ with respect to some norm $ \norm{ \cdot } $ in $ R^{n_x} $ into $ M $ disjoint sets $ \bar{Q}_1,...,\bar{Q}_M \subset \mathbb{R}^{n_x} $ each equipped with a point $ \bar{q}_i \in \bar{Q}_i $ called the centroid of $ \bar{Q}_i $. Such a quantizer associates any vector $ x \in \bar{Q}_i $ with its quantized value $ \bar{q}_i $. The quantizer $ \mathscr{Q} $ is said
		to be $ r- $contracted $ (r = 1, 2,...) $ for the system \eqref{lti1} if	
		\begin{align*}
			A^r (\bar{Q}_i-\bar{q}_i) \subset \rho_{\mathscr{Q}} \mathbf{B}_1, \quad \forall i=1, \dots ,M,
		\end{align*}		
		where $ \rho_{\mathscr{Q}} \in (0,1) $ is the {\em contraction rate}.
	\end{defi}
	Note that the quantizer is linked to the system dynamics in \eqref{lti1} through matrix $ A $. The following lemma relates the number of levels to matrix $A$ in \eqref{lti1}.
	\begin{lem} [{\cite[Ch. 3]{matveev2009estimation}}] \label{lem:lvl}
		For any $  r- $contracted quantizer, the following inequality holds
		\begin{align*}
			M > |\det A|^r = 2^{r h_{lin}}.
		\end{align*}
		Moreover, there exists an $  r- $contracted quantizer such that 
		\begin{align*}
			M \leq 2^{n_x-1} \big[\varpi(r)^{n_x} + 1\big] \, 2^{r h_{lin}} ,
		\end{align*}
		where $ \varpi(r) $ is a polynomial function of $ r $.
	\end{lem}
	\begin{rema}
		Lemma \ref{lem:lvl} shows that for a large $ r $ the number of levels for the coarsest $  r- $contracted quantizer is roughly $ M\approx 2^{rh_{lin}} $.
	\end{rema}		
	In what follows, we study the bounded state estimation problem over communication channels with feedback.
	\subsection{State estimation in the presence of channel feedback} \label{sub:est}
	{In this subsection, the main result for the estimation problem over a causal channel that has errorless feedback from its output is presented.
	The structure of this problem is shown in Fig. \ref{fig:estf}. The plant is an LTI  system and is affected by a bounded process noise, $ V_t $. For the estimation problem, we assume {that the plant is not controlled, i.e.,} $  U_t = 0, \, \forall t \in \mathbb{N} $.}
	
	{{The encoder maps the plant state sequence $ X_{1:t} $ and previous channel outputs $ Y_{1:t-1} $ to the channel input, i.e., 
	\begin{align*}
		Q_t =\mu^e_t(X_{1:t},Y_{1:t-1})\in \mathcal{X},
	\end{align*}
	where $ \mu^e $ is an encoder operator. Each symbol $ Q_t $ is then transmitted over the channel.} The received symbols $ Y_{1:t} \in \mathcal{Y} $ are decoded at the decoder and a causal prediction $ \hat{X}_{t+1} $ of $ X_{t+1} $ is produced by means of another operator $ \eta^e $ as 
	\begin{align*}
	\hat{X}_{t+1}=\eta^e_t(Y_{1:t}) \in \mathbb{R}^{n_x},
\end{align*}
		with the estimation initialized at $\hat{X}_{1}=0$. The sequence of pairs $ \{(\mu^e_t,\eta^e_t)\}_{t\geq 1} $ is called a {\em coder-estimator}.
	
	We have the following theorem.
	\begin{thm} [Bounded estimation with feedback] \label{thm:finitem}
		Consider an LTI system \eqref{lti1} satisfying conditions A1--A4 and $  U_t~=~0$, $\forall t \in \mathbb{N} $. Assume that outputs are coded and estimated via a causal channel (Definition \ref{def:causc})  having errorless feedback from channel output to the encoder
		{with positive zero-error feedback capacity, i.e., $C_{0f}>0$. If the estimation error is uniformly bounded then
			\begin{align}
				C_{0f} \geq h_{lin}. \label{estwf}
			\end{align}
			Conversely, if $C_{0f} > h_{lin}$, then there exists a coder-estimator that keeps the estimation error uniformly bounded.}
	\end{thm}
	\begin{proof}
		See Appendix \ref{app:finitem}.
	\end{proof}
	\begin{rema} \textcolor{white}{.}
		
		\begin{itemize}
			\item Theorem \ref{thm:finitem} states that uniformly reliable estimation is possible if the zero-error feedback capacity of the channel exceeds the rate at which the system generates uncertainty.
			\item If $ A $ has no eigenvalue with magnitude larger than $ 1 $ as opposed to A4, then the right-hand side of \eqref{estwf} is zero and the inequality already holds for any positive capacity.
		\end{itemize}
	\end{rema}
	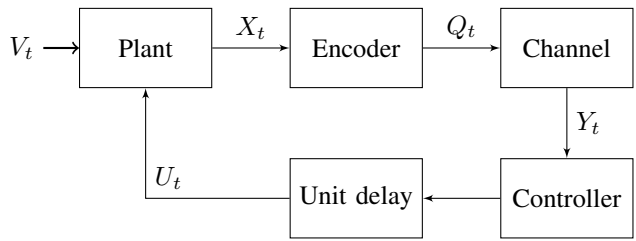
\begin{figure}[t]
		\centering
		\scalebox{1}{\begin{tikzpicture}[auto, node distance=1cm,>=latex']
				\node [block2, pin={[pinstyle]left: $ V_t $ }] (plant) { Plant };
				\node [block1, right of=plant, node distance=2.8cm] (encoder) { Encoder };
				\node [block2, right of=encoder, node distance=2.8cm] (channel) { Channel };
				\node [block1, below of=channel, node distance=2cm] (decoder) { Controller} ;
				\node [block2, below of=encoder, node distance=2cm] (delay) {Unit delay };
				
				\draw [draw,->] (plant) -- node {$ X_t $} (encoder);
				\draw [->] (encoder) -- node {$ Q_t $} (channel);
				\draw [->] (channel) -- node {$ Y_t $} (decoder);
				\draw [->] (decoder) -- node [name=u] {}(delay);
				\draw [->] (delay) -| node [above right] {$ U_t $} (plant);
		\end{tikzpicture}}
		\caption{ Stabilization problem via a communication channel. }
		\label{fig:ctrl}
	\end{figure}
	\subsection{Stabilization}\label{sub:ctrl}
	{The structure of the stabilization problem is shown in Fig. \ref{fig:ctrl}. We assume the states of the linear system are fully observed.		
		The encoder maps the system state to the channel input, i.e.,
		\begin{align*}
			Q_t=\mu^c_{t}(X_{1:t}) \in \mathcal{X},\quad t\in \mathbb{N}.
		\end{align*}
		Each symbol $ Q_t $ is then transmitted over the channel. The received symbol $ Y_t \in \mathcal{Y} $ is decoded at the controller (decoder) and a causal control signal $ U_t$ is produced, i.e.,
		\begin{align*}
			U_t &= \eta^c_{t}(Y_{1:t}) \in \mathbb{R}^{n_u}.
		\end{align*}
		The pair $ (\mu^c,\eta^c) $ is called a {\em coder-controller}.}
	\begin{thm} [Bounded stabilization] \label{thm:stab}
		{Consider the LTI system \eqref{lti1} that satisfies conditions A1–-A5 with states controlled via a causal channel [Definition \ref{def:causc}] (without feedback). The LTI system and channel can be initialized at any time $ t_0+1 \in \mathbb{N} $ such that it belongs to a nonempty ball $ \mathbf{B}_l  \subseteq \mathbb{R}^{n_x}$ with known radius $ l $. If the closed-loop system state is uniformly bounded then
			\begin{align}
				C_{0f} \geq h_{lin}. \label{stabdis}
			\end{align}
			Conversely, if $C_{0f} > h_{lin}$, then there exists a coder-controller that keeps the states of the system uniformly bounded.}
	\end{thm}
	
	\begin{proof}
		See Appendix \ref{app:stab}.
	\end{proof}
	\begin{rema}
		\textcolor{white}{.}
		
		\begin{itemize}
			\item This theorem signifies that, in fact, the zero-error capacity with complete feedback defines the conditions for stability, {even though} there is no explicit feedback from the channel output to the encoder. 
			\item This result holds for a general class of channel models with memory in which the channel error patterns may not be just i.i.d. and are correlated.
			\item The proof is based on creating an implicit feedback loop using the control signal, hence, the appearance of $ C_{0f} $ rather than $ C_0 $. If the communication channel were to incorporate noiseless feedback, the construction of the implicit feedback {is} not required, and the same result {would} follow. Therefore, having communication feedback does not improve the stability condition.
		\end{itemize}
	\end{rema}
	{In what follows, we consider a special channel model and derive its zero-error feedback capacity. We will return to the stabilization problem in Section \ref{sec:stabofsan}.}
	\section{FSANC Model and Zero-error Capacity Results} \label{sec:model}
	In this section, we turn our attention to finite-state additive noise channels as a subclass of FSCs (Definition \ref{def:fschannel}). We first define these channels and then the main results for the zero-error capacity with(out) feedback are given. Some additional properties for the FSANC are discussed in Section \ref{sec:apc0} followed with some examples in Section \ref{sec:examps}. {In Section \ref{sec:stabofsan}, the zero-error capacity results for FSANCs are combined with the bounded stabilization condition of Theorem \ref{thm:stab} and a tight condition for stabilization over such channels is given as a corollary.}
	
	We use a graph to describe the state evolution in the following notion. 	
	\begin{defi}[Finite-state machine] \label{def:fsm}
		A {\em finite-state machine} is defined as {a} directed graph $\mathscr{G}=(\mathcal{S},\mathcal{E})$, where the vertex set $ \mathcal{S}=\{0,1,\dots,|\mathcal{S}|-1\} $ denotes {the} {\em states} of the machine, and the edge set $\mathcal{E}\subseteq\mathcal{S}\times \mathcal{S}$ denotes possible transitions between two states. Each edge takes a value $ z \in \mathcal{Z} $ that corresponds to the output of the process. The possible edges outgoing from each state depend only on the current state and
		\begin{equation*}
			S_{i+1}  = \mathfrak{f}(S_{i},\varPhi_i), \quad i \in \mathbb{N}, 
		\end{equation*}
		where $ \varPhi_i \in \mathcal{N}_s$ is mutually unrelated to $ (\varPhi_{1:i-1},S_1) $, and  $ \mathfrak{f}:\mathcal{S} \times \mathcal{N}_s \to \mathcal{S}  $. 
		
		Furthermore, a finite-state machine (or its graph) is strongly connected if every state is reachable from every other state, that is, there is a directed path in the graph from every state to every other state \cite{lind1995introduction}.
	\end{defi}
	In other words, {the} next state is mutually unrelated {to} past states given the current state, forming a Markov uncertainty chain, i.e.,
	\begin{align*}
		\llbracket S_{i+1}|s_{1:i} \rrbracket =\llbracket S_{i+1}|s_i \rrbracket, \quad i \in \mathbb{N} .
	\end{align*}
	\begin{rema}
		In a stochastic setting, processes described by a finite-state machine are {\em topologically Markov} \cite[Ch.1]{adler1979topological}, \cite[Ch.2]{lind1995introduction}, which is weaker than the standard Markov property.
		In a topological Markov chain, the (probability 1) set of allowed next-states given past and present states depends only on the present state. However, the conditional probability of the next state may depend on past as well as present states, violating the stochastic Markov property. An example of such a system is studied in Example \ref{estat} in {Section} \ref{sec:examps}.
	\end{rema} 
	
	In this section, the following channel is studied.
	\begin{defi}[Finite-state additive noise channels]
		\label{def:fanchannel}	
		A discrete channel with common input, noise and output $ q $-ary alphabet $ \mathcal{X} $ is called {\em finite-state additive noise} if its output at time $ i \in \mathbb{N} $ is obtained by
		\begin{align*}
			Y_i = X_i \oplus Z_i,
		\end{align*}
		where the correlated additive noise $ Z_i $ is governed by a state process $ (S_i) $ on a finite-state machine such that each outgoing edge from a state $ s_i $ corresponds to different values $ z_i $ of the noise. Thus, there are at most $ q $ outgoing edges from each state. {Let the initial condition $ S_1 \in \mathcal{S} $ be an uncertain variable that is not known in advance.}
		We assume the finite-state machine is strongly connected and that $ Z_i= \mathfrak{h}(S_{i},\varPhi_i) $ where $ \varPhi_i \in \mathcal{N}_s$ is mutually unrelated to $ (\varPhi_{1:i-1},S_1) $, and  $ \mathfrak{h}:\mathcal{S} \times \mathcal{N}_s \to \mathcal{S}  $. In other words,
		$ \llbracket Z_i | s_{1:i},  x_{1:i} \rrbracket = \llbracket Z_i|s_i \rrbracket $. 
	\end{defi}
	{Here, because each outgoing edge from any state on $\mathscr{G}$ corresponds to different noise values, there is a one-to-one correspondence between state and noise sequences, and therefore, $ \varPhi_i $ is also shared between state updates and noise output mappings.}
	
	Figure \ref{fig:mealy} shows a noise process which defines a channel that no two consecutive errors can happen. For example, the transition at time $ i $ from state $ S_i=0 $ to itself corresponds to $ Z_i=0 $. Moreover, $ Z_i=1 $ leads to the transition ending in state $ S_{i+1}=1 $ (state at next time step). Note that, in {$ S_i=1 $}, the noise can only take $ Z_i=0 $ and transits to $ S_{i+1}=0 $. 
	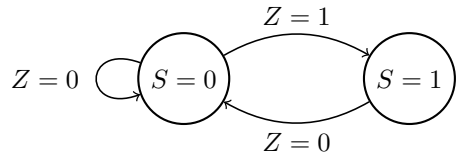
\begin{figure}[t]
		\centering
		\begin{tikzpicture}
			[->, auto, semithick, node distance=3cm]
			\tikzstyle{every state}=[fill=white,draw=black,thick,text=black,scale=1]
			\node[state]   (S1)       {$ S=0 $};
			\node[state]   (S2)[right of=S1]   {$ S=1 $};
			\path
			(S1) edge[out=160, in=200,looseness=5] node [left=0.1cm] {$ Z=0 $}(S1)
			edge[ bend left] node [above] {$ Z=1 $}  (S2)
			(S2) edge[ bend left]  node [below] {$ Z=0 $}  (S1);
		\end{tikzpicture}
		\caption{State transition diagram of a noise process in a channel at which no two consecutive errors can happen in the channel.}
		\label{fig:mealy}
	\end{figure}
	\begin{defi}[Coupled graph] \label{def:dg}
		The {\em coupled graph} of a finite-state machine (with labeled graph $ \mathscr{G} $) is defined as a labeled directed graph\footnote{ This product is called {\em tensor product},  or the {\em Kronecker  product} \cite[Ch. 4]{hammack2011handbook}.}  $\mathscr{G}_c=\mathscr{G}\times \mathscr{G}$, such that it has vertex set $ \mathcal{V}=\mathcal{S}\times \mathcal{S} $ and has an edge from node $ u=(i,j) \in \mathcal{V} $ to $ v=(k,m) \in \mathcal{V}  $ {iff} there are edges from $ S=i $ to $ S=k $ (with a label value $ E_{ik} $) and from $ S=j $ to $ S=m $ (with a label value $ E_{jm} $) in $ \mathscr{G} $, each edge has a label equals to $E_{ik} \ominus E_{jm}$.
	\end{defi}
	The coupled graph of the finite-state machine in Fig. 1 is demonstrated in Fig. 2.
	
	Before presenting the main results of this section, we give some preliminaries from symbolic dynamics.
	In symbolic dynamics, topological entropy is defined as the asymptotic growth rate of the number of possible  state sequences. Define the state {\em transition matrix} $ \mathcal{A} \in \{0,1\}^{|\mathcal{S}|\times |\mathcal{S}|}$ such that the $(s,s')$th entry $\mathcal{A}_{s,s'}$ equals 1 if the state of the channel can transition from $s$ to $s'$, and equals 0 otherwise. For a finite-state machine with an {\em irreducible} transition matrix $\mathcal{A}$, the topological entropy $ h $ is known to coincide with $\log\lambda$, where $\lambda$ is the {\em Perron value} of $\mathcal{A}$~\cite{lind1995introduction}.\footnote{The unique largest real eigenvalue of an irreducible, square, and non-negative matrix is called the Perron value \cite{lind1995introduction}.} This is essentially due to the fact that the number of the paths from state $ S=i $ to $ S=j $ in $ n $ steps is the $ (i+1,j+1) $-th element of $ \mathcal{A}^n$, which grows at the rate of $\lambda^n$ for large $n$.
	
	\subsection{Zero-error capacities of {FSANC}s} \label{sec:mrc0}
	Now, we give a condition on when zero-error capacity is zero, with or without feedback.
	\begin{thm}\label{thm:c0f0}
		The zero-error capacity with(out) feedback $ C_{0f} \,$(resp. $ C_0 $) of a {FSANC} (Definition \ref{def:fanchannel}) having finite-state machine (Definition \ref{def:fsm}) graph  $\mathscr{G}=(\mathcal{S},\mathcal{E}) $  is zero, {iff} $ \forall \, d_{1:n} \in \mathcal{X}^{n} , n \in \mathbb{N} $, there exists a walk on the {\em coupled graph} (Definition \ref{def:dg}) of $ \mathscr{G} $ with the label sequence $ d_{1:n} $. 
	\end{thm}
	\begin{proof}
		See Appendix \ref{app:c0f0}.
	\end{proof}
	\begin{cor}
		For {FSANC}s $ C_0 = 0 $ {iff} $ C_{0f} = 0 $.
	\end{cor}
	
	The following lemma relates the channel output size to the topological entropy of the channel, i.e., $ h_{ch}=\log \lambda $.
	\begin{lem}\label{lem:out}
		For a {FSANC} with irreducible adjacency matrix, there exist positive constants $ \alpha $ and  $ \beta $ such that, for any input sequence $ x_{1:n} \in \mathcal{X}^n $, the number of all possible outputs
		\begin{align}
			\alpha \lambda^n \leq \big|\mathcal{Y}(s_1,x_{1:n}) \big|=\big|\mathcal{Z}(s_1,n) \big| \leq  \beta  \lambda^n , \label{outlam}
		\end{align}
		where $\lambda $ is the Perron value of the adjacency matrix. Moreover, $ \mathcal{Y}(s_1,x_{1:n}) $ and $ \mathcal{Z}(s_1,n) $ are the possible output and noise values for a given initial state $ s_1 $ and input sequence $ x_{1:n} $.
	\end{lem}
	\begin{proof}
		See Appendix \ref{app:out}.
	\end{proof}
	We now relate the zero-error capacities of the channel to the noise process topological entropy.
	
	\begin{thm}\label{thm:zero-error}
		The zero-error feedback capacity of the {FSANC} (Definition \ref{def:fanchannel}) with topological entropy $h_{ch}$ of the noise process is either zero or 
		\begin{align}
			C_{0f} &= \log q-h_{ch}.\label{c0f}
		\end{align}
	\end{thm}
	Moreover, in \cite{saberi2020bounded}, we have shown that the zero-error capacity (without feedback) is lower bounded by
	\begin{align}
		C_{0} &\geq \log q - 2h_{ch}.\label{c0l}
	\end{align}
	\begin{proof}
		In the following, we show the achievability of \eqref{c0f} {in case $ C_{0f} >0 $}. The converse for \eqref{c0f} is shown in Appendix \ref{app:zero-error}.
		
		The conditions on when $ C_{0f}=0 $ is given in Theorem \ref{thm:c0f0}. Here, we consider $ C_{0f}>0 $. 
		\begin{figure}[t]
			\centering
			\begin{tikzpicture}
				[->, auto, semithick, node distance=3cm]
				\tikzstyle{every state}=[fill=white,draw=black,thick,text=black,scale=1]
				\node[state]   (S1)       {$ (0,0) $};
				\node[state]   (S2)[right of=S1]   {$ (1,1) $};
				\node[state]   (S3)[below of=S1]   {$ (1,0) $};
				\node[state]   (S4)[right of=S3]   {$ (0,1) $};
				\path
				(S1) 	 edge[out=100, in=150,looseness=5] node [left=0.1cm] {$ 0 $}(S1)
				(S1)	 edge[bend left] node [above] {$ 0 $} (S2)
				(S1.300) edge[bend right] node [above] {$ 1 $}  (S4.140)
				(S1.250) edge[bend right] node [left] {$ 1 $}  (S3.110)
				(S2.200) edge[bend left]  node [above] {$ 0 $}  (S1.350)
				(S4)	 edge[bend left] node [above] {$ 1 $}  (S3)
				(S3.80)  edge[bend right]  node [left] {$ 0 $}  (S1.280)
				(S3.30)  edge[bend left] node [below] {$ 1 $}  (S4.160)
				(S4) 	 edge[bend right]  node [left] {$ 0 $}  (S1.330);
			\end{tikzpicture}
			\caption{Coupled graph of the noise process in Fig. \ref{fig:mealy}.  }
			\label{fig:cg}
		\end{figure}
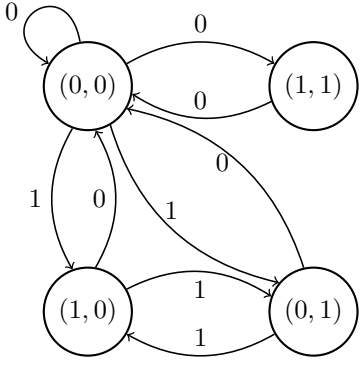		
		
		A coding method {that achieves \eqref{c0f} is proposed}. Consider a code of length $ n $ such that the first $ k (<n) $ symbols are the data to be transmitted, and the rest of $ n-k $ symbols serve as parity check symbols. 
		
		{Let $\mathcal{Y}(k):=\cup_{s_1} \mathcal{Y}(s_1,k)$. From Lemma \ref{lem:out}, we have
			\begin{align*}
				\big|\mathcal{Y}(k)\big|= \big|\cup_{s_1} \mathcal{Z}(s_1,k)\big| 
		\end{align*}}
		possible output sequences, which is bounded {for any $ s_1 \in \mathcal{S} $} as follows
		\begin{align}
			\alpha \lambda^k \leq \big|\mathcal{Z}(s_1,k)\big| \leq \big|\mathcal{Y}(k)\big| \leq |\mathcal{S}| \beta \lambda^k. \label{outie}
		\end{align}
		{The transmitter having output sequence  $y_{1:k-1}$	sends} the receiver which output pattern (e.g., a message from $\{1,\dots, |\mathcal{Y}(k)|\} $ ) was received using the $ n-k $ parity check symbols. In other words, since the output size and the noise set are equal for {a} given input, i.e., \eqref{yvsize}, using the parity check symbols the decoder is informed of the noise sequence for the first $ k $ symbols. Since for {a given noise sequence }\eqref{adnoise} is bijective, the decoder obtains $ x_{1:k} $.  
		
		Assume that the transmitter sends the parity check symbols with a rate slightly below the zero-error feedback capacity, i.e., $ R=C_{0f}-\delta $, where $ \delta >0 $ is arbitrarily small. {By Lemmas \ref{lem:delc0f} and \ref{lem:canc} such a code with rate $ R=C_{0f}-\delta $ exits.}
		Therefore, 
		\begin{align*}
			C_{0f}-\delta &= \frac{\log \big|\mathcal{Y}(k)\big|}{n-k}.
		\end{align*}
		Using the upper bound on size of the output in \eqref{outie}, we obtain
		\begin{align*}
			C_{0f}-\delta \leq \frac{\log \big(|\mathcal{S}| \beta\lambda^k \big) }{n-k}.
		\end{align*}
		Rearranging the inequality gives
		\begin{align*}
			k &\geq \frac{(C_{0f}-\delta)n-\log(|\mathcal{S}|\beta)}{(C_{0f}-\delta +\log \lambda)}.
		\end{align*}
		Considering the fact that the total rate of coding is upper-bounded by $ C_{0f} $, we have
		\begin{align*}
			C_{0f} &\geq \frac{k}{n}\log q \\
			&\geq \frac{C_{0f}-\delta-\frac{\log(|\mathcal{S}|\beta)}{n}}{C_{0f}-\delta +\log \lambda} \log q.
		\end{align*}
		Rearranging gives the following.	
		\begin{align*}
			C_{0f} & \geq  \log q-\log \lambda  -\delta \big(1-\frac{1}{C_{0f}}\big)\log q -\frac{\log\big(|\mathcal{S}|\beta\big)}{nC_{0f}} \log q.
		\end{align*}
		By choosing $ \delta $ small and making $ n $ large, the last two terms disappear and this concludes the achievability proof.
	\end{proof}
	\begin{rema}
		\textcolor{white}{.}
		
		\begin{itemize}
			\item The zero-error feedback capacity has a similar representation to the ordinary feedback capacity, i.e., \mbox{$ C_{f} = \log q-\mathcal{H}_{ch} $} in \cite{alajaji1995feedback}, but with the stochastic noise entropy rate  $ \mathcal{H}_{ch} $ replaced with the topological entropy $h_{ch}$. {Generally, $ C_f \geq C_{0f} $ and therefore, $ \mathcal{H}_{ch} \leq h_{ch} $. The interpretation is that $h_{ch}$ is the asymptotic rate of growth in number of all {possible} trajectories (worst-case) for any given input and $ \mathcal{H}_{ch}$ is the asymptotic rate of growth in the number of {\em typical} trajectories given the input. The latter is not bigger than the topological entropy.}
			
			\item The topological entropy can be viewed as the rate at which the noise dynamics generate uncertainty. Intuitively, this uncertainty cannot increase the channel capacity, which explains why it appears as a negative term on the right-hand side of \eqref{c0f} and \eqref{c0l}. Moreover, the sum of zero-error feedback capacity and the topological entropy is always equal to $ \log q $, meaning that if the noise uncertainty is increased, the feedback capacity decreases by the same amount.
			
			\item The result of \eqref{c0f} is an explicit closed-form solution, which is a notable departure from the iterative, dynamic programming solution {derived for {FSC}s with available state information} in \cite{zhao2010zero}.
			
			\item {Inequality} \eqref{c0l} implies that when $ h_{ch} <\frac{1}{2}\log q  $, the zero-error capacity is non-zero ($ C_0>0 $) and so  \mbox{$ C_{0f}=\log q- h_{ch}>0 $}.
			
			\item {The essential ingredients in {the} proof of Theorem \ref{thm:zero-error} are the bijective property between the noise and output sequences, given the channel input, and {the unrelatedness of the} channel input and channel noise. Therefore, the additive nature of the channel is not necessary. This suggests that it is possible to extend these results to channels beyond FSANCs. }
			
			\item Following Definition \ref{def:fanchannel}, the channel states are not assumed to be Markov, just topologically Markov. Thus in the case of having a probabilistic structure, the transition probabilities in the finite-state machine can be time-varying or dependent on the previous states. In other words, as long as the graphical structure has not changed, the result is valid.
		\end{itemize}
	\end{rema}
\begin{rema}\label{rem:vp}
	The topological entropy $ h_{ch} $ and the process entropy rate $ \mathcal{H}_{ch} $ are related by the {\em variational principle} which states that $ h_{ch} = \sup_{P \in \mathcal{I}} H_{ch}(P)$.  Here, $  \mathcal{I}$ is the set of all probability measures on the finite-state machine (with graph $\mathscr{G}$) which are invariant under all automorphisms\footnote{An automorphism of $\mathscr{G}$ is a bijection of the vertex set which preserves adjacencies.} 
	of $ \mathscr{G} $ \cite{parry1964intrinsic}. As a consequence of Theorem 4, it follows that for finite-state additive noise channels with positive $  C_{0f} $, 
	\begin{align}
 	 C_{0f} &= \log q - h_{ch} \nonumber\\
				&=\log q - \sup_{P \in \mathcal{I}} H_{ch}(P) \label{varp}\\
				&=  \inf_{P\in \mathcal{I}} C_f(P). \nonumber
	\end{align}
In other words, Theorem \ref{thm:zero-error} leads to a variational principle connecting $ C_{0f}  $ and $ C_f $ for finite-state additive noise channels.
For channels outside this class, it is unlikely that  a variational principle for feedback capacity holds. 
\end{rema}
	\subsection{Some properties of {FSANC}s}\label{sec:apc0}
	In what follows, some properties of {FSANC}s including a condition on when $ C_0=0 $ and how to bound the zero-error capacities based on known cases are given. 
	\begin{figure}[t]
		\centering
		\begin{tikzpicture}
			[->, auto, semithick, node distance=3.5cm]
			\tikzstyle{every state}=[fill=white,draw=black,thick,text=black,scale=0.8]
			\node[state]   (S1)       {$ S=0 $};
			\node[state]   (S2)[right of=S1]   {$ S=1 $};
			\node[state]   (S3)[right of=S2]   {$ S=2 $};
			\path
			(S1) edge[out=160, in=200,looseness=5] node [above=0.25] {$ Z=0 $}(S1)
			edge[bend left] node [above] {$ Z=1 $}  (S2)
			(S2) edge[bend left]  node [above=0.1] {$ Z=0 $}  (S1)
			edge[bend left]  node [above] {$ Z=1 $}  (S3)
			(S3) edge[bend left]  node [above] {$ Z=0 $}  (S1.300);
		\end{tikzpicture}
		\caption{\small{A finite-state machine describing the transition of a noise process in a channel at which no more than two consecutive errors can happen.}}
		\label{fig:mealy2}
	\end{figure}
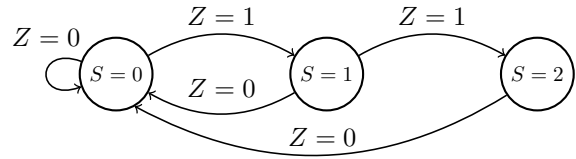
	Let $ \mathcal{Z}(n) $ be the set of all possible noise sequences starting from any initial condition. Define 
	\begin{align*}
		\mathcal{P}(n):=\big\{z_{1:n} \ominus z'_{1:n}| z_{1:n},z'_{1:n} \in \mathcal{Z}(n) \big\}.
	\end{align*}
	We now give the following lemma.
	\begin{lem}\label{lem:zcz}
		$ C_0 = 0 $ iff $\forall n \in \mathbb{N}, \, \mathcal{P}(n)=\mathcal{X}^n $.
	\end{lem}
	\begin{proof}
		See Appendix \ref{app:zcz}.
	\end{proof}
	\begin{defi}\label{subg}
		Let $ \mathscr{G}_A = (\mathcal{V}_A, \mathcal{E}_A) $ and $ \mathscr{G}_B = (\mathcal{V}_B, \mathcal{E}_B) $. The graph $ \mathscr{G}_A $ is a subgraph of $ \mathscr{G}_B $ if both $ \mathcal{V}_A \subseteq \mathcal{V}_B $ and $ \mathcal{E}_A \subseteq \mathcal{E}_B $.
	\end{defi}
	\begin{lem} \label{lem:win}
		Let $\varpi_A(n,v_0)$ be a walk (i.e., {a sequence of $ n $ edge labels}) on $ \mathscr{G}_A $ starting from vertex $ v_0 \in \mathcal{V}_A $. If $ \mathscr{G}_A $ is a subgraph of $ \mathscr{G}_B $ then for any walk on $ \mathscr{G}_A $ there is a walk on $ \mathscr{G}_B $ such that $\varpi_A(n,v_0)=\varpi_B(n,v_0)$.
	\end{lem}
	\begin{proof}
		See Appendix \ref{app:win}.
	\end{proof}
	\begin{prop}\label{prop:extended0f0}
		Let $\mathscr{G}_A$ and $\mathscr{G}_B$ be graphs for noise processes of two {FSANC}s $ A $ and $ B $ with zero-error capacities $ C^A_{0} $ and $ C^B_{0} $, respectively. If $\mathscr{G}_A$ is an induced subgraph of $\mathscr{G}_B$ then $ C^A_0 \geq C^B_0 $. 
	\end{prop}
	\begin{proof}
		See Appendix \ref{app:extended0f0}.
	\end{proof}
	In the remainder of the section, a few examples of {FSANC}s are discussed.
	
	\subsection{{FSANC} examples} \label{sec:examps}
	Here, we provide some examples to compute $ C_{0f} $ explicitly. Examples \ref{2stat} and \ref{3stat} consider channels with isolated and limited runs of errors. A memoryless channel is investigated in Example \ref{memless} and compared with Shannon's result. In Example \ref{estat}, we consider a Gilbert-Elliott channel. {Finally, in Example \ref{swc}, we discuss a worst-case channel model, referred to as sliding-window, in which the number of errors in every sliding window is upper bounded by a known number \cite{saberi2020bounded}.} Moreover, for Examples \ref{2stat} and \ref{3stat}, we investigate the minimum value of ordinary feedback capacity $ C_f $ over the transition probabilities and observe how far is this natural upper bound from the zero-error feedback capacity.\smallskip
	
	\begin{ex}\label{2stat}
		{A commonly used constraint in optical and magnetic storage systems is the Run Length Limited (RLL) constraint \cite{Zehavi88}. We consider an example channel of this type in Fig.~\ref{fig:mealy} in which no two consecutive 1's can happen. This constraint is also known as $(1,\infty)$-RLL constraint.}
		\begin{prop}\label{prop:ex0f0}
			The zero-error capacity with or without feedback of the channel in Fig. \ref{fig:mealy} with a binary alphabet $ (q=2) $ is zero.
		\end{prop}
		\begin{proof}
			See Appendix \ref{app:ex0f0}.
		\end{proof}
		Whilst if $ q \geq 3 $, the channel of Fig. \ref{fig:mealy} has nonzero zero-error capacities, since a simple repetition code of length two is a zero-error code with $ C_0>0 $. From Theorem \ref{thm:zero-error}, we then have
		\begin{align*}
			C_{0f}=\log q-\log(\frac{1+\sqrt{5}}{2})
		\end{align*}
		bit/use where $ \frac{1+\sqrt{5}}{2} $ is known as the {\em golden ratio}.
		
		Moreover, assuming Markovianity with the transition probability $ P(S_{i+1}=1|S_i=0)=p $, using the result of \cite{alajaji1995feedback} the ordinary feedback capacity for $q\geq 2$  is 
		\begin{align*}
			C_f (p)= \log q-\frac{H(p)}{1+p},
		\end{align*}
		where $ H(\cdot) $ is the binary entropy function. It can be shown that 
		\begin{align*}
			\inf_{p\in (0,1)} C_f (p) =\log q-\log(\frac{1+\sqrt{5}}{2}).
		\end{align*} 
	This is equal to $ C_{0f} $ when $q\geq 3$, verifying that the variational principle for feedback capacity discussed in Remark \ref{rem:vp} holds. Note this principle does not hold when $q=2$ (in this case, Proposition \ref{prop:ex0f0} implies that $ C_{0f} = 0 $, but the RHS above is positive).
	\end{ex} \smallskip
	
	\begin{ex}\label{3stat}
		The example of Fig.~\ref{fig:mealy2} represents a channel with no more than two consecutive errors with adjacency matrix
		\begin{align*}
			\mathcal{A}=\begin{bmatrix}
				1&1&0\\
				1&0&1\\
				1&0&0
			\end{bmatrix}.
		\end{align*}
		If $ q = 2 $, then $ C_0=C_{0f}=0 $. This follows from Propositions \ref{prop:extended0f0} and \ref{prop:ex0f0}, as the graph in Fig.~\ref{fig:mealy} is a subgraph of Fig.~\ref{fig:mealy2}. Whereas, if $ q =3 $, then $ C_{0f} = 0.7058$. 
		If the channel states are Markov with transition probabilities $ P(S_{i+1}=1|S_i=0)=p $ and $ P(S_{i+1}=2|S_i=1)=r $, it can be shown that for $ q=2,3, $
		\begin{align*}
			C_f (p,r) &= \log q - \frac{1}{1+p+rp} H(p) -\frac{p}{1+p+rp} H(p).
		\end{align*}
	It can further be shown that $ \inf_{p,r\in (0,1)}C_f (p,r) = C_{0f}$ when $q=3$, verifying the variational principle for feedback capacity discussed in Remark \ref{rem:vp}. However if $q=2$ the variational principle fails, since the infimum is positive while $C_{0f} = 0$.
	\end{ex}\smallskip
	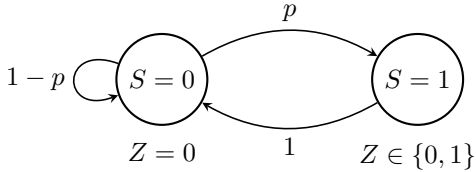
\begin{figure}[t]
		\centering
		\begin{tikzpicture}
			[->, >=stealth, auto, semithick, node distance=3.4cm]
			\tikzstyle{every state}=[fill=white,draw=black,thick,text=black,scale=1]
			\node[state] (A1)   {$ S=0 $};
			\node[] (state0)[below=0.1cm of A1] {$ Z=0 $};
			\node[state]   (A2)[right of=A1]   {$S=1 $};
			\node[] (state1)[below=0.1cm of A2] {$  Z \in \{0,1\} $};
			\path
			(A1) edge[out=160, in=200,looseness=5] node [left] {$ 1-p $} (A1)
			edge [bend left] node [above] {$ p $}  (A2)
			(A2) edge[bend left]  node [below] {$ 1$}  (A1);
		\end{tikzpicture}
		\caption{Markov chain for channel states in Example \ref{estat}.}
		\label{fig:estat}
	\end{figure}
	
	\begin{ex}\label{memless}
		Consider a $ q $-ary memoryless additive noise channel such that each input $ x_i \in \mathcal{X} $ can get mapped to $ y_i=x_i \oplus z_i \in~\mathcal{X}, \, z_i \in \mathcal{Z} \subseteq \mathcal{X},\,\forall i \in \mathbb{N}$. We assume $ |\mathcal{X}|>2|\mathcal{Z}| $, hence, $ C_{0f}>0 $. This channel is memoryless and a generalization of binary symmetric channel. At each time, the noise can take $ |\mathcal{Z}| $ number of values. Thus, $ h_{ch} = \log |\mathcal{Z}|$. From \eqref{c0f}, we obtain
		\begin{align}
			C_{0f}=\log \frac{|\mathcal{X}|}{|\mathcal{Z}|}.\label{mlc0f}
		\end{align}
		
		Similarly, we can show that Shannon's formula in \cite{shannon1956zero} gives the same result (see Appendix \ref{app:shan}).  Lov{\'a}sz derived an upper bound for zero-error capacity in \cite{lovasz1979shannon} by which, for this channel with $ \mathcal{Z}=\{0,1\} $ and odd $|\mathcal{X}| $, it yields the following bound.
		\begin{align}
			C_0 \leq \log \frac{|\mathcal{X}|}{1+1/\cos (\pi/|\mathcal{X}|)}.
		\end{align}
		Therefore, for such channels, $ C_{0f} > C_0 $.
	\end{ex}
	In what follows, we study an example of Gilbert-Elliott channel which is {an FSC} with two states $ \mathcal{S}= \{0, 1\} $,
	the state $ S=0 $ corresponding to the ``good'' state and state $ S=1 $, to the ``bad'' state \cite{mushkin1989capacity, goldsmith1996capacity}. The channel has equal input and output alphabets and the probability law
	\begin{align*}
		P(y,s'|x,s)=P(y|x,s)P(s'|s),
	\end{align*}
	where $ P(s'|s) $ is the transition probability from state $ s \in \mathcal{S} $ to state $ s' \in \mathcal{S} $, and $ P(y|x,s) $ is the conditional distribution of a DMC given current state $ s $. Usually, the DMCs are considered to be binary symmetric channels such that the channel corresponding to the bad state has higher crossover probability than the DMC of the good state. Obviously, in this case, the overall channel has $ C_0=C_{0f}=0 $. To investigate a non-trivial case, a channel with alphabet size 5 is considered.
	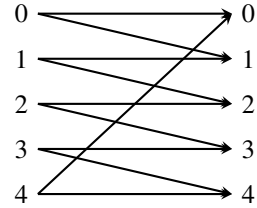
\begin{figure}
		\centering
		\begin{tikzpicture}[->, >=stealth, auto, thick, node distance=0.6cm]
			\tikzstyle{every state}=[fill=white,thick,text=black,scale=.6]
			\node[]    (S1)  				{0};
			\node[]    (S2)[below of=S1]   {1};
			\node[]    (S3)[below of=S2]   {2};
			\node[]    (S4)[below of=S3]   {3};
			\node[]    (S5)[below of=S4]   {4};
			\node[]    (S6)[right of=S1,right=2.2cm]   {0};
			\node[]    (S7)[below of=S6]   {1};
			\node[]    (S8)[below of=S7]   {2};
			\node[]    (S9)[below of=S8]   {3};
			\node[]    (S10)[below of=S9]   {4};
			\path
			(S1.0) edge   (S6)
			edge    (S7.180)
			(S2.0) edge  (S7)
			edge  (S8.180)
			(S3.0) edge   (S8)
			edge    (S9.180)
			(S4.0) edge  (S9)
			edge  (S10.180)
			(S5.0) edge   (S10)
			edge    (S6.180);
		\end{tikzpicture}
		\caption{Pentagon channel.}
		\label{fig:pen}
	\end{figure}
	\begin{ex}\label{estat} Consider a Gilbert-Elliott channel with an input alphabet of size $ q=5 $ and two states (Fig.~\ref{fig:estat}). When the state $ S_i=0 $ the channel is error-free, i.e., $ P(Z_i\neq 0|S_i=0)= 0 $ and when state $ S_i=1 $ it acts like a noisy type-writer channel (Fig.~\ref{fig:pen}) which is also known as the Pentagon channel \cite{shannon1956zero}. In this state, the probability of error for any input symbol is $ P(Z_i=1|S_i=1)= r $ and thus the probability of error-free transmission is $ P(Z_i=0|S_i=1)=1-r $. Figure~\ref{fig:estat} shows this channel's state transition diagram. However, this channel does not fit Definition \ref{def:fanchannel}, because outgoing edges are not associated with unique noise values. This reflects the fact that the noise process is a hidden Markov model, not a Markov chain, and the same state sequence can yield multiple noise sequences.  
		
		Nonetheless, in the following, we show an equivalent representation of this channel compatible with Definition \ref{def:fanchannel}. The resultant model (shown in Fig.~\ref{fig:eq2}) is a state machine that produces the same set of noise sequences, where the edges define the noise values in each transmission.
		
		Note that if the channel is in state $ S_i=0 $, the noise can only take value $ Z_i=0 $, but in state $ S_i=1 $, the noise $ Z_i\in \{0,1\} $, thus $ Z_i\in \{0,1\} $ at all times. In the sequel, we show that 
		\begin{align}
			P\big(Z_{i+1}=1|Z_i=1,z_{1:i-1}\big)&=0,\label{zz11}\\
			P\big(Z_{i+1}=0|Z_i=0,z_{1:i-1}\big)&>0,\label{zz00}\\
			P\big(Z_{i+1}=1|Z_i=0,z_{1:i-1}\big)&>0,\label{zz10}
		\end{align}
		whenever the conditioning sequence of $ Z_i=j,Z_{1:i-1}=z_{1:i-1}, j\in \{0,1\} $ occurs with non-zero probability. Therefore, irrespective of past noises the state machine shown in Fig.~\ref{fig:eq2} can produce all noise sequences that occur with nonzero probability. It should be stressed that this noise process may not be a stochastic Markov chain, however, it is a topological Markov chain \cite[Ch.2]{lind1995introduction}. 
		First, note by inspection of Fig.~\ref{fig:estat} that the noise process has zero probability of taking value $ 1 $ twice in a row. Thus 
		$ P\big(Z_{i+1}=1, Z_i =1, z_{1:i-1}\big) = 0 $.
		{Using Bayes rule, \eqref{zz11} then follows}.
		
		Next we show \eqref{zz00}-\eqref{zz10}. Let $z_{1:i-1}$ be any past noise sequence such that $P\big(Z_i=0, z_{1:i-1}\big) >0$. Therefore, $\exists s_{1:i} $ such that 
		\begin{align}
		\begin{split}P\big(Z_i=0,z_{1:i-1},s_{1:i}\big)&=\\
			P&\big(Z_i=0|s_i\big)P(z_{1:i-1},s_{1:i})>0.
		\end{split}\label{con0}
	\end{align}
		From Fig.~\ref{fig:estat}, $ P\big(Z_{i+1}=0,Z_i=0|s_i=j\big)>0, j\in \{0,1\} $. Thus
		\begin{align*}
			\begin{split}
				P\big(Z_{i+1}=0,Z_i=0,z_{1:i-1},s_{1:i}\big) &=\\ P\big(Z_{i+1}=0,Z_i&=0|s_i\big)P(z_{1:i-1},s_{1:i})>0,
			\end{split}
		\end{align*}
		since the second factor on the {right-hand side} is positive, by \eqref{con0}. Therefore, $ P(Z_{i+1}=1, Z_i=0, z_{1:i-1}) > 0 $, and \eqref{zz00} holds. Now, we show \eqref{zz10}. If $ S_i=0 $, it can be shown from Fig.~\ref{fig:estat} and the noise probabilities that 
		\begin{align}
			P\big(Z_{i+1}=1,Z_i=0|S_i=0\big)&=rp>0. \label{z10rp}
		\end{align}
		Therefore, 
		\begin{align*}
			P\big(&Z_{i+1}=1,Z_i=0,z_{1:i-1}\big) \\
			&\geq P\big(Z_{i+1}=1,Z_i=0,z_{1:i-1},S_i=0, s_{1:i-1}\big)\\
			&=P\big(Z_{i+1}=1,Z_i=0|S_i=0\big) P\big(S_i=0,z_{1:i-1},s_{1:i-1}\big)\\
			&=rp\,P\big(S_i=0,z_{1:i-1},s_{1:i-1}\big)> \, 0.
		\end{align*}
		Note from Fig.~\ref{fig:estat} that $ P\big(S_i=0|s_{i-1}\big)>0 $. Thus,	
		\begin{align*}
			P\big(S_i=0,z_{1:i-1},s_{1:i-1}\big)&=P\big(S_i=0|s_{i-1}\big)P(z_{1:i-1},s_{1:i-1})\\
			& > \, 0.
		\end{align*}
		Consequently, \eqref{zz11}-\eqref{zz10} hold yielding the state machine in Fig.~\ref{fig:eq2}. Note that, $ \hat{S}_i=0 $ corresponds to $ Z_i=0 $ and  $ \hat{S}_i=1 $, to $ Z_i=1 $.
		\begin{figure}[t]
			\centering
			\begin{tikzpicture}
				[->, auto, semithick, node distance=3.4cm]
				\tikzstyle{every state}=[fill=white,draw=black,thick,text=black,scale=1]
				\node[state]   (S1)       {$ \hat{S}=0 $};
				\node[state]   (S2)[right of=S1]   {$ \hat{S}=1 $};
				\path
				(S1) edge[out=160, in=200,looseness=5] node [left=0.1cm] {$ Z=0 $} (S1)
				edge[bend left]  node [above] {$ Z=1 $}  (S2)
				(S2) edge[bend left]  node [below] {$Z=0 $}  (S1);
			\end{tikzpicture}
			\caption{State machine generating the noise sequence of Example \ref{estat}.}
			\label{fig:eq2}
		\end{figure}
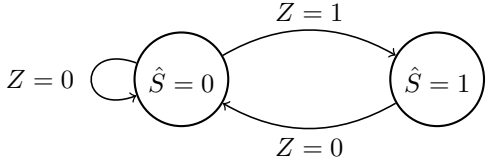
		
		Now, we can use the results of Theorem \ref{thm:zero-error}, to obtain	
		\begin{align*}
			C_0 &\geq \log5-2\log\bigg(\frac{1+\sqrt{5}}{2}\bigg), \\
			C_{0f} &= \log5-\log\bigg(\frac{1+\sqrt{5}}{2}\bigg).
		\end{align*}
		This shows that the zero-error feedback capacity of some channels with different structure than Definition \ref{def:fanchannel}, such as time-varying state transmissions (non-homogeneous Markov chains) and even transitions that depend on previous transmissions can be explicitly obtained.		\end{ex}\smallskip
	\begin{figure}[t]
		\centering
		\scalebox{0.9}{\begin{tikzpicture}[->, >=stealth', auto, semithick, node distance=3cm]
				\tikzstyle{every state}=[fill=white,draw=black,thick,text=black,scale=0.9]
				\node[state]   (S1)       {$  S=000 $};
				\node[state]    (S2)[left of=S1]   {$ S=001 $};
				\node[state]    (S3)[below of=S2]   {$ S=010 $};
				\node[state]    (S4)[right of=S3]   {$ S=100 $};
				\path
				(S1) edge[out=30,in=70,looseness=8] (S1)
				edge[bend right]  node [above] {$Z=1 $}(S2)
				(S2) edge[bend right] node [left] {$Z=0 $} (S3)
				(S3) edge[bend right] node [above] {$Z=0 $} (S4)
				(S4) edge[bend right]  node [right] {$Z=0 $} (S1)
				(S4) edge[]  node [right] {$Z=1 $} (S2);
		\end{tikzpicture}}
		\caption{State machine generating the noise sequence of a sample ($3,1$) sliding-window channel (Example \ref{swc}).}
		\label{fig:swc}
	\end{figure}
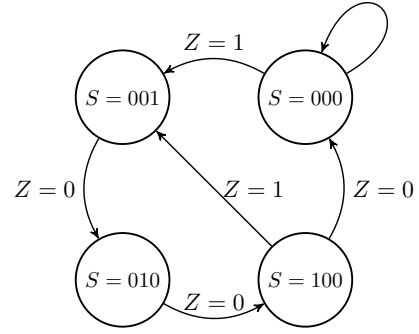
	\begin{ex}\label{swc}
		A binary $(w,d)$ sliding-window channel, has at most $ d $ errors in each sliding window of length $ w $ \cite{saberi2020bounded}. We define the state as a binary word of length $w$, in which $0$ indicates no error and $1$ label the erroneous symbol swaps that can occur. Therefore, there are $ \sum_{k=0}^d{w \choose k} $ states for the noise process based on the position and number of errors. Equivalently, by writing the channel input-output relationship as $Y_i=X_i \oplus Z_i$, the current channel state is equivalent to $Z^{i-1}_{i-w}\in\mathbb{Z}_2^w$, with at most $d$ nonzero entries.  This is not the most compact state representation; however, for a given input sequence, it yields a one-to-one relationship between the state and output sequences. Using the results of Theorem \ref{thm:zero-error}, the zero-error feedback capacity of these types of channels can be derived. As an example consider a binary ($w=3,d=1$) sliding-window channel with noise state process shown in Fig. \ref{fig:swc} has $ C_{0f}=0.449 $. Note that in this case, the lower bound on the zero-error capacity gives a negative value $ -0.103 $, which demonstrates its conservativeness.
	\end{ex}
	
	\subsection{Stabilization over {FSANC}s} \label{sec:stabofsan}
	{In this subsection}, by combining the results above, {a tight} condition is obtained for uniformly bounded stabilization problem via {FSANC}s, involving the topological entropies of the linear system and the channel. 
	\begin{cor} [{Small-Entropy Theorem}] \label{thm:fsan_st} 
		Consider an LTI system \eqref{lti1} satisfying conditions A1--A4. Assume that $ (A,B) $ is stabilizabile and the measurements are coded and transmitted via a {FSANC} (Definition \ref{def:fanchannel})
		with alphabet set size $ q $ and topological entropy $h_{ch}$. Then uniform stabilization can be achieved if
		\begin{align}
			h_{lin} + h_{ch}<\log q.\label{twoent}
		\end{align}
		Conversely, if $ h_{lin} + h_{ch}>\log q $, no encoder-controller can keep the states of the linear system uniformly bounded.
	\end{cor}
	
	\begin{proof} This follows from Theorems \ref{thm:zero-error} and \ref{thm:stab}.  \end{proof}
	\begin{rema}
		{Inequality} \eqref{twoent} involves the topological entropies of both the linear system and the channel. If their sum, which can be regarded as a total rate of uncertainty generation, is less than the worst-case rate at which symbols can be transported without error across the channel, then uniform stabilization is possible. This is similar to the well-known {``Small-Gain Theorem"} in control \cite[Ch. 3]{desoer2009feedback}, albeit for uncertainty rather than gain.
	\end{rema}  
	\section{Conclusion} \label{sec:conclusions}
	We introduced a formula for computing the zero-error feedback capacity of a class of additive noise channels without state information at the decoder and encoder. This reveals a close connection between the topological entropy of the underlying noise process and zero-error communication.
	
	We showed that the necessary and sufficient condition for uniformly bounded stabilization over communication channels is that the zero-error feedback capacity is larger than the topological entropy of the plant. This result gives a tight condition for the control over communication channels problem without having any statistical information about uncertainties. 
	
	Furthermore, combining two major results of this paper, i.e., the zero-error feedback capacity of the {FSANC} and bounded control condition, reveals a ``{Small-Entropy Theorem}" for stabilization over {FSANC} with alphabet size $ q $ which states that uniformly bounded stabilization is possible {if} the sum of topological entropies of the plant and channel is smaller than $ \log q $.
	
	Future work includes extending {the zero-error capacity results to continuous alphabet channels with correlated but bounded noise. Exploring the trade-offs between the performance of the controlled linear system and the channel capacity is another research direction.}
	
	\appendices
	\section{Proof of Lemma \ref{lem:canc}}\label{app:fsccaus}
	{ By Definition \ref{def:fschannel}, we have
		\begin{align}
			S_{i+1} &= \mathfrak{f}(\mathfrak{f}\big(S_{i-1}, X_{i-1},\varPhi_{i-1}\big),X_{i},\varPhi_{i})\nonumber\\
			& = \mathfrak{f}( \dots \mathfrak{f}\big(S_1, X_1,\varPhi_1\big), \dots,X_{i},\varPhi_{i})\nonumber\\
			&=:\mathfrak{f}^i(S_1,X_{1:i},\varPhi_{1:i}). \label{fupi}
		\end{align}
		Substituting above in \eqref{fsyc} yields
		\begin{align}
			Y_i &= \mathfrak{h} \big(\mathfrak{f}^{i-1} (S_1,X_{1:i-1},\varPhi_{1:i-1}),X_i, \varPsi_i\big) \nonumber \\
			&=: \bar{g}_i(S_1,X_{1:i},\varPhi_{1:i-1},\varPsi_i) \label{outcauss}\\
			&=: g_i(X_{1:i},Z_{1:i}), \nonumber
	\end{align}}
	where $ \bar{g}_i $ and $ g_i $ are defined based on the functional relationship of the arguments to the channel output $ Y_i $, and $ Z_1:=(S_1,\varPhi_1,\varPsi_1),\, Z_{i}:=(\varPhi_{i},\varPsi_{i}), i\geq 2 $. {This also shows that the effect of history prior to time 1 is summarized in $ S_1 $ as it appears in $ Z_1 $.} By Definition \ref{def:fschannel}, $ Z_i $ is mutually unrelated to $ Z_{1:i-1} $. This shows that the output is only a function of past and current input and an unrelated noise which satisfies the condition in Definition \ref{def:causc}.
	
	\section{Proof of Proposition \ref{prop:fsc0f}}\label{app:fsc0f}
	We first show that under the encoding scheme in \eqref{zef} and having $ t_0 \in \mathbb{N}\cup\{0\} $, the possible output sequences in \eqref{fsyc} from a subset of the possible outputs when $ t_0=0 $ given message $ m \in \mathcal{M}$, i.e., 
	\begin{align}
		\llbracket Y_{t_0+[1:n]}|m \rrbracket & \subseteq \llbracket Y_{1:n} |m \rrbracket,\quad \forall t_0 \in \mathbb{N}\cup\{0\}. \label{t0-t}
	\end{align} 
	We derive \eqref{t0-t} by induction such that $ \forall i\in \{1, \dots , n\} $,
	\begin{align*}
		\llbracket  Y_{t_0+[1:i]} |m \rrbracket &\subseteq \llbracket Y_{1:i}|m \rrbracket. 
	\end{align*} 
	{\em Base step}: at $ i=1 $ and $ t_0=0 $, {since the prior state information are not available for both the encoder and decoder,} all the states in $ \mathcal{S} $ are possible. By \eqref{zef}, $ x_{1}=f_1(m) $, hence,
	\begin{align} 
		\llbracket Y_{1}|m \rrbracket &= \llbracket Y_{1}|m,x_{1} \rrbracket = \bigcup_{s_{1} \in \mathcal{S}} \llbracket Y_{1}|x_{1},s_{1} \rrbracket. \label{y1m}
	\end{align} 
	The last equality in \eqref{y1m} follows from \eqref{outun} as the output only depends on the current input and state. Similarly, starting at time $ t_0+1 > 1 $, by \eqref{zef}, $ x_{t_0+1}=x_1=f_1(m) $, and
	\begin{align} 
		\llbracket Y_{t_0+1}|m \rrbracket &= \llbracket Y_{t_0+1}|m,x_{t_0+1} \rrbracket\\
		&= \bigcup_{s_{t_0+1} \in \mathcal{S}_{t_0+1}} \llbracket Y_{t_0+1}|x_{1},s_{t_0+1} \rrbracket, \label{yt01m}
	\end{align} 
	where $ \mathcal{S}_{t_0+1}:=\bigcup_{s_1\in \mathcal{S}}\llbracket S_{t_0+1}|s_1 \rrbracket \subseteq \mathcal{S}$. The communication starts at time $ t_0+1 $ and the encoder and the decoder has no access to the previous transmissions. However, the set of possible states at time  $ t_0+1 $ can be smaller. This depends on the mapping in \eqref{fssc} whose image may not include all the states in $ \mathcal{S} $. For example if the mapping in \eqref{fssc} is not surjective, then there exists a particular state $ s^* \in \mathcal{S}_1= \mathcal{S}$ such that $ s^* \notin \mathcal{S}_{2} $. This argument holds for the next updates of the state in \eqref{fssc}.\footnote{ One may consider {an FSC} represented by a directed graph that a node with no incoming edge and therefore can not be visited more than once, i.e., only if it is the initial state.} Therefore, $ \mathcal{S}_{t_0+1} \subseteq \mathcal{S}_1=\mathcal{S}$. From \eqref{y1m} and \eqref{yt01m}, we obtain
	\begin{align*}
		\llbracket  Y_{t_0+1} |m \rrbracket &\subseteq \llbracket Y_{1}|m \rrbracket. 
	\end{align*} 
	{\em Inductive step}: assume the following holds
	\begin{align}
		\llbracket Y_{t_0+[1:i-1]}|m \rrbracket &\subseteq \llbracket Y_{1:i-1}|m \rrbracket. \label{indstep}
	\end{align}
	For any output sequences $ \mathbf{y} \in \llbracket Y_{t_0+[1:i-1]}|m \rrbracket$, by \eqref{zef}, $ \mathbf{x}=x_{t_0+[1:i]}=x_{1:i}=f_{1:i}(m,\mathbf{y}) $.
	Since the {FSC} in \eqref{fssc}-\eqref{fsyc} is time-invariant starting from a state $ s $ and input $ x_{1:i} $ same set of outputs can be produced at  $ i $ starting from any time, i.e.,
	\begin{align}
		\llbracket Y_{i}|x_{1:i}=\mathbf{x},S_1=s \rrbracket &= \llbracket Y_{t_0+i}|x_{t_0+[1:i]}=\mathbf{x}, S_{t_0+1}=s \rrbracket. \label{yinits}
	\end{align}
	We have
	\begin{align}
		\llbracket Y_{i}|m,Y_{1:i-1}=\mathbf{y} \rrbracket &= \llbracket Y_{i}|m, x_{1:i}=\mathbf{x}, \mathbf{y} \rrbracket  \nonumber\\
		&=\bigcup_{s \in \bar{\mathcal{S}}_1} \llbracket Y_{i}|\mathbf{x},\mathbf{y},s\rrbracket, \label{yiyki}
	\end{align} 
	where $\bar{\mathcal{S}}_1:=  \llbracket S_{1}|x_{1:i}=\mathbf{x},Y_{1:i-1}=\mathbf{y} \rrbracket $.
	Similarly, having $ Y_{t_0+[1:i-1]}=\mathbf{y} $, $ \mathbf{x}=x_{t_0+[1:i]} $, and from \eqref{yinits},
	\begin{align}
		\llbracket Y_{t_0+i}|m,\mathbf{y} \rrbracket &= \llbracket Y_{t_0+i}|m, x_{t_0+[1:i]}, \mathbf{y}\rrbracket \nonumber \\
		&=\bigcup_{s\in \bar{\mathcal{S}}_{t_0+1}} \llbracket Y_{t_0+i}|\mathbf{x},\mathbf{y},s \rrbracket, \label{yiyki2}
	\end{align} 
	where $\bar{\mathcal{S}}_{t_0+1}:=  \llbracket S_{t_0+1}|x_{t_0+[1:i]}=\mathbf{x}, Y_{t_0+[1:i-1]}=\mathbf{y} \rrbracket $.
	\begin{lem}\label{lem:inits}
		\begin{align*}
			\bar{\mathcal{S}}_{t_0+1} \subseteq \bar{\mathcal{S}}_1.
		\end{align*}
	\end{lem}
	\begin{proof}
		We show that any $ s \in  \bar{\mathcal{S}}_{t_0+1}$ also belongs to $ \bar{\mathcal{S}}_{1} $.
		From \eqref{outcauss}, any output sequence is a function of the past and current channel input and noise sequences as well as the initial state. Choose any admissible $ s \in  \bar{\mathcal{S}}_{t_0+1} \subseteq \mathcal{S}_{t_0+1} $, i.e., $ \exists \, \mathbf{p} \in  \llbracket\varPhi_{t_0+[1:i-2]},\varPsi_{t_0+[1:i-1]}\rrbracket $ that produces $ Y_{t_0+[1:i-1]}=\mathbf{y} $.
		Here, $ \tilde{\mathbf{x}}:= x_{t_0+[1:i-1]} $. In other words,
		\begin{align}
			Y_{t_0+[1:i-1]}=\bar{g}_{1:i-1} \big(s,\tilde{\mathbf{x}},\mathbf{p}\big)=\mathbf{y}.\label{outputtili-1}
		\end{align}
		Because the channel input at each time is a function of message $ m $ and previous outputs, \eqref{outputtili-1} can be simplified by defining function $ \tilde{g}_i:\mathcal{S} \times \mathcal{M} \times  \big( \mathcal{N}_s^{i-1} \times \mathcal{N}_o^{i} \big) \to \mathcal{Y}$, i.e.,
		\begin{align}
			Y_{t_0+[1:i-1]}=\tilde{g}_{1:i-1} \big(s,m,\mathbf{p}\big).\label{outputsimp}
		\end{align}
		Now, choose $ s_1=s \in \mathcal{S}_1 $ (this is valid as shown in the base step, $ \mathcal{S}_{t_0+1} \subseteq \mathcal{S}_1$). Also, choose $ (\phi_{1:i-2},\psi_{1:i-1})=\mathbf{p} $ (valid by Definition \ref{def:fschannel}), and $ x_{1:i-1}=\tilde{\mathbf{x}} $. The same output $ Y_{1:i-1}=\mathbf{y} $ is produced by \eqref{outcauss} which according to \eqref{indstep} is in the set of possible outputs, i.e.,
		\begin{align*}
			Y_{[1:i-1]}=\tilde{g}_{1:i-1} \big(s,m,\mathbf{p}\big)=\bar{g}_{1:i-1} \big(s,\tilde{\mathbf{x}},\mathbf{p}\big)=\mathbf{y}.
		\end{align*}
		This yields $ s \in  \bar{\mathcal{S}}_{1} $ and therefore, $\bar{\mathcal{S}}_{t_0+1} \subseteq \mathcal{S}_{1} $.	
	\end{proof}
	
	By Lemma \ref{lem:inits}, \eqref{yiyki}, and \eqref{yiyki2},
	\begin{align*}
		\llbracket Y_{t_0+i}|m,Y_{t_0+[1:i-1]}=\mathbf{y} \rrbracket & \subseteq \llbracket Y_{i}|m,Y_{1:i-1}=\mathbf{y} \rrbracket.
	\end{align*}
	Considering all the possible outputs of $ y_{1:i} \in \llbracket Y_{1:i}|m \rrbracket  $, we obtain 
	\begin{align}
		\llbracket Y_{t_0+[1:i]}|m \rrbracket &=\bigcup_{\mathbf{y} \in \llbracket Y_{t_0+[1:i-1]}|m \rrbracket}  \{\mathbf{y}\} \times \llbracket  Y_{t_0+i}|m,\mathbf{y}\rrbracket \nonumber\\
		& \subseteq \bigcup_{\mathbf{y} \in \llbracket Y_{1:i-1}|m \rrbracket} \{\mathbf{y}\} \times \llbracket Y_{i}|m,\mathbf{y}\rrbracket \nonumber \\
		&=\llbracket Y_{1:i}|m \rrbracket, \label{t0to0}
	\end{align}
	where $ \times $ denotes Cartesian product. Hence, \eqref{t0to0} proves the inductive step and so \eqref{t0-t} holds. Because any output set corresponding to a transmission starting at $ t_0+1 $ is a subset of the output set with $ t_0=0 $, thus, any zero-error code of blocklength $ n $ yields distinguishable messages for any shift, $ t_0 $ in transmissions. In other words, any zero-error feedback code is a uniform zero-error feedback code for {an FSC}. 
	
	\section{Proof of Theorem \ref{thm:finitem} (state estimation in presence of channel feedback)}\label{app:finitem}
	We give the proof of the Theorem \ref{thm:finitem} in the sequel. 
	\subsubsection{Converse}
	\begin{lem} \label{lem:finitem}
		Suppose that states of the system in \eqref{lti1} are coded and estimated with an arbitrary encoder-decoder pair via a channel with feedback such that
		\begin{align}
			C_{0f} < h_{lin}. \label{estnec}
		\end{align}
		{Then, an admissible sequence $ \{X_t \in \mathbb{R}^{m} \}_{t \in \mathbb{N}} $ and a channel realization exist for which the estimation error is unbounded, i.e., 
		\begin{align}
			\limsup_{t\to \infty }\|X_t - \hat{X}_t\|=\infty. \label{unbes}
		\end{align}}
	\end{lem}
	
	\begin{proof} 
		The plan of the proof is as follows.
		
		\begin{itemize}
			\item For proof by contradiction, we assume that \eqref{unbes} fails to be true, i.e., $ \exists \phi $ such that $ \limsup_{t\to \infty }\|X_t - \hat{X}_t\| < \phi $;
			\item Under this condition, we construct a uniform zero-error feedback code that achieves $ R \approx h_{lin} $;
			\item By the definition of the zero-error feedback capacity, this contradicts \eqref{estnec}. Hence, this contradiction proves that \eqref{unbes} must be correct.
		\end{itemize}	
		Suppose $ Q_t=\mu^e_t(X_{1:t},Y_{1:t-1}) \in \mathcal{X}, t \in \mathbb{N} $ be the channel's input where $ \mu^e $ is the encoder operator. Each symbol $ Q_t $ is then transmitted over the channel. The received symbols $ Y_{1:t} \in \mathcal{Y} $ are decoded and a causal prediction $ \hat{X}_{t+1} $ of $ X_{t+1} $ is produced by means of another operator $ \eta^e $ as $\hat{X}_{t+1}=~\eta^e_t(Y_{1:t}) \in \mathbb{R}^{n_x}, \,\hat{X}_1=0$. {Let the estimation error be denoted by $ E_t := X_t-\hat{X}_t$ .}
		
		Assume a coder-estimator achieves uniform bounded estimation error. By change of coordinates, it can be assumed that $ A $ matrix is in {\em real Jordan canonical form} which consists of $ \varrho $ square blocks on its diagonal, with the $ j $-th block $ A(j) \in \mathbb{R}^{n_j\times n_j} ,\, j=1,\dots,\varrho$. Let $ X_t(j),\hat{X}_t(j)=\eta^e_t\big(Y_{1:t-1}(j)\big) \in \mathbb{R}^{n_j} $ and so on, be the corresponding $ j $-th component.
		
		Let $ \kappa \in \{1,\dots,n_x\} $ denote the number of eigenvalues with magnitude larger than $ 1 $ including repeated values. From now on, we will only consider the unstable subsystem, as the stable part plays no role in the analysis. We use the same line of reasoning in \cite{nair2013nonstochastic} to construct a subset of the LTI plant's initial state that lead to non-overlapping outputs. Considering that the initial point belongs to a $ l $-ball $ \mathbf{B}_l \subseteq \mathbb{R}^n$, by picking $
		\epsilon \in (0, 1-\max_{i:\lambda_i|>1} |\lambda_i|^{-1})$,
		arbitrary $ \nu \in \mathbb{N} $, and dividing the interval $ [-l,l] $ on the $ i $-th axis into 
		\begin{align}
			k_i:=\lfloor|(1-\epsilon)\lambda_i|^{\nu-1}\rfloor, \, i \in \{1,\dots,\kappa\} \label{kk}
		\end{align}
		equal subintervals of length $ 2l/k_i $. Let $ p_i(s),\, s=\{1,\dots,k_i \}$ denote the midpoints of the subintervals and inside each subinterval construct an interval $ \mathbf{I}_i(s) $ centered at $ p_i(s) $ with a shorter length of $ l/k_i $. A hypercuboid family is defined as below 
		\begin{align}
			\mathscr{I}&=\bigg\{\bigg( \prod_{i=1}^{\kappa} \mathbf{I}_i(s_i)\bigg) :  s_i \in\{1,\dots,k_i\}, i \in \{1,\dots,\kappa\}\bigg\}, \label{hyper}
		\end{align}
		in which any two hypercuboids are separated by a distance of $ l/k_i $ along the $ i $-th axis for each $ i \in \{1,\dots,\kappa\}$. Now, consider an initial point with range 
		\begin{align}
			\llbracket X_1\rrbracket =\cup_{\mathbf{L}\in \mathscr{I}} \mathbf{L} \subset \mathbf{B}_l \subset \mathbb{R}^{\kappa}. \label{xsub}
		\end{align}
		Let diam($ \cdot $) denote the set diameter under the {${\infty} $ or max-}norm and given the received sequence $ y_{1:t-1} $. {First, we consider the case that the process noise is zero, i.e., $ V_t = 0, \forall t \in \mathbb{N} $. This assumption will be relaxed later. We have}
		\begin{align}
			\text{diam}  &\llbracket {E_t(j)| V_{1:t}=0} \rrbracket \nonumber \\
			\geq& \text{ diam} \llbracket {E_t(j)}| y_{1:t-1}{,V_{1:t}=0} \rrbracket \label{condi}\\
			=&\text{ diam} \llbracket X_t(j)-\eta^e_t\big(y_{1:t-1}(j)\big)| y_{1:t-1} {,V_{1:t}=0}\rrbracket \nonumber\\
			=&  \text{ diam} \llbracket  [A(j)]^{t-1}X_1(j)+ \Theta(j,V_{1:t-1})| y_{1:t-1}{,V_{1:t}=0}\rrbracket\label{condi2}\\
			\geq&  \text{ diam} \llbracket  [A(j)]^{t-1}X_1(j)| y_{1:t-1},V_{1:t}=0\rrbracket \nonumber\\
			\geq& \sup_{u,v \in \llbracket  X_1(j)| y_{1:t-1}{,V_{1:t}=0}\rrbracket}\frac{\norm{[A(j)]^{t-1}(u-v)}_2}{\sqrt{n_x}} \nonumber\\
			\geq& \sup_{u,v \in \llbracket  X_1(j)| y_{1:t-1}{,V_{1:t}=0}\rrbracket}\frac{\sigma_{min}([A(j)]^{t-1})\norm{u-v}_2}{\sqrt{n_x}} \nonumber\\
			=& \sigma_{min}([A(j)]^{t-1})\frac{\text{diam} \llbracket X_1(j)| y_{1:t-1}{,V_{1:t}=0}\rrbracket}{\sqrt{n_x}}\label{diamf},
		\end{align}
		where $\Theta (j,V_{1:t-1})= \sum_{i=1}^{t-1} [A(j)]^{t-1-i}V_i(j) $ and $\sigma_{min}(\cdot) $ denotes smallest singular value. \eqref{condi} holds since conditioning reduces the range \cite{nair2013nonstochastic}. Note that \eqref{condi2} follows from the fact that translating does not change the range.
		Using Yamamoto identity \cite[Thm. 3.3.21]{roger1994topics}, $ \exists t_{\epsilon} \in \mathbb{N} $ such that $ \forall t\geq t_{\epsilon} $ the following holds
		\begin{align}
			\sigma_{min}([A(j)]^t) \geq \big(1-\frac{\epsilon}{2}\big)^t \big|\lambda_{min}(A(j))\big|^t, \;j=1,\dots,p. \label{yamam}
		\end{align}
		By bounded state estimation error hypothesis $ \exists \phi>0  $, such that
		\begin{align*}
			\phi &\, \geq \sup \big\llbracket {\norm{E_t}|V_{1:t}=0}  \big\rrbracket \\
			&\, \geq \sup \big \llbracket {\norm{E_t(j)}|V_{1:t}=0} \big \rrbracket\\ 
			&\, \geq 0.5 \,\text{diam}\sup \llbracket {\norm{E_t(j)}|V_{1:t}=0}  \rrbracket  \\
			& \stackrel{\eqref{diamf}}{\geq} \big((1-\frac{\epsilon}{2})|\lambda_{min}(A(j))|\big)^{t-1}\frac{\text{diam} \llbracket X_1(j)| y_{1:t-1}{,V_{1:t}=0} \rrbracket}{2\sqrt{n_x}} 
		\end{align*}
		Now, we show that for large enough $ \nu $ and the {hypercube family $ \mathscr{I} $ \eqref{hyper} is an $ \llbracket X_1| y_{1:\nu}, V_{1:\nu}=0\rrbracket $-overlap isolated partition of $ \llbracket X_1\rrbracket $. By contradiction, suppose that $ \exists \mathbf{L}\in \mathscr{I} $ that is overlap connected in $ \llbracket X_1| y_{1:\nu}, V_{1:\nu}=0\rrbracket $ with another hypercube in $ \mathscr{I} $.} Thus there exists a conditional range $ \llbracket X_1| y_{1:\nu-1}{,V_{1:\nu}=0}\rrbracket $ containing both a point $ u\in \mathbf{L} $ and a point $ v $ in $ \mathbf{L}' \in \mathscr{I}\backslash \mathbf{L}$. Henceforth, $ \forall \nu \geq t_{\epsilon} $,
		\begin{align*}
			\norm{u-v}\leq& \text{ diam}\llbracket X_1(j)| y_{1:\nu-1}{,V_{1:\nu}=0}\rrbracket\\
			\leq& \frac{2\sqrt{n_x}\phi}{\big((1-{\epsilon /2})\big|\lambda_{min}(A_j)\big|\big)^{\nu-1}} \;j=\{1,\dots,p\}.
		\end{align*}
		Notice that by construction any two hypercuboid in $ \mathscr{I} $ are separated by a distance of $ l/k_i $, which implies
		\begin{align*}
			\norm{u_j-v_j} &\geq \frac{l}{k_i}\\ &=\frac{l}{\lfloor(1-\epsilon)|\lambda_i|\rfloor^{\nu-1}} \\
			&\geq \frac{l}{\big|(1-\epsilon)\lambda_{min}(A_j)\big|^{\nu-1}}
		\end{align*}
		The right hand side of this equation would exceed the right hand side of \eqref{yamam}, when $ \nu  $ is large enough that 
		\begin{align*}
			\bigg( \frac{1-\epsilon/2}{1-\epsilon}\bigg)^{\nu-1}> 2\frac{\sqrt{n_x}\phi}{l},
		\end{align*}
		yielding a contradiction.	
		Therefore, for sufficiently large $ \nu $, no two sets of $ \mathscr{I} $ {are $ \llbracket X_1| y_{1:\nu-1},V_{1:\nu}=0\rrbracket $-overlap connected. 
			The number of hypercubes in $ \mathscr{I} $ satisfy}
		\begin{align}
			|\mathscr{I}|&= \prod_{i=1}^{\kappa} k_i\\
			&=\prod_{i=1}^{\kappa}\lfloor |(1-\epsilon)\lambda_i|^{\nu-1} \rfloor\\	
			&> \prod_{i=1}^{\kappa} 0.5 \big|(1-\epsilon)\lambda_i\big|^{\nu-1} \label{half}\\
			&=2^{-\kappa}(1-\epsilon)^{\kappa(\nu-1)}\bigg|\prod_{i=1}^{\kappa}\lambda_i\bigg|^{\nu-1},\label{sizep}
		\end{align}
		where \eqref{half} holds since \mbox{$ \lfloor x \rfloor > x/2, \, \forall x > 1 $}. Here, \mbox{$ x=(1-\epsilon)\lambda_i >1 $} holds by choosing 
		\begin{align}
			\epsilon < 1-\frac{1}{\min_i \lambda_i}.
		\end{align}
		\begin{rema}\label{rema:zede}
			For any two initial conditions $ x \in \mathbf{L}, x'\in \mathbf{L}',\, \mathbf{L}\neq \mathbf{L}',\, \mathbf{L}, \mathbf{L}' \in \mathscr{I}$, we have 
			\begin{align}
				\llbracket Y_{1:\nu-1}|x{,V_{1:\nu}=0} \rrbracket \cap  \llbracket Y_{1:\nu-1}|x'{,V_{1:\nu}=0} \rrbracket= \emptyset,
			\end{align}
			otherwise, $ x $ and $ x' $ belong to a single partition which contradicts with $ \mathbf{L}\neq \mathbf{L}' $.	
		\end{rema}	
		{Based on Remark \ref{rema:zede}, by choosing any two points in distinct hypercubes in $ \mathscr{I} $ (e.g., the center point of each hypercube), the corresponding outputs do not overlap. Here, the $ \llbracket X_1| y_{1:\nu-1},V_{1:\nu}=0 \rrbracket$-overlap isolated partition can be used to construct a zero-error code for the channel. That it also can be used as a zero-error code. We denote this zero-error code with $ \mathcal{F}_x $ which satisfies $ |\mathcal{F}_x|=|\mathscr{H}| $.}
		
		Now we show that this leads to a contradiction with respect to the zero-error feedback capacity of the communication channel. 
		
		We consider a trajectory for the plant in \eqref{lti1} {with $ U_t=0 $ and} $ V_t= 0, \, \forall t \in \mathbb{N} $. Therefore, 
		\begin{align}
			X_t=A^{t-1}X_1,\quad t\in \mathbb{N}\, .
		\end{align} 	
		We construct a zero-error feedback code for the channel with a message set size of $ |\mathscr{I}| $ using the same encoder-decoder for the state estimation problem. Therefore every message $m \in {1, \dots,  |\mathscr{I}| }$  is assigned to the center point of the hypercube $x^c_m \in \mathbf{L} \in \mathscr{I}$ using the following function. Note that the encoder has a unit delay feedback from the channel's output.
		\begin{align}
			q_t:=\mu^e_t\big(x_{1:t},y_{1:t-1}\big),\quad t=1, \dots,\nu.
		\end{align}
		Here, $ x_{1:t}=\big(A^{i-1}x^c_m\big)_{i=1}^t $ and the goal is to transmit $x^c_m \in \mathbf{L} $ across the channel. At the decoder, each received symbol sequence is mapped to the corresponding message $m$. By virtue of Remark \ref{rema:zede}, we must have unambiguous (zero-error) decoding for sufficiently large $ \nu $. 
		The encoder can also be represented by
		\begin{align}
			q_t=\mu^e_t\bigg(\big(A^{i-1}x^c_m\big)_{i=1}^t,y_{1:t-1}\bigg) = f_t( m, y_{1:t-1}), \label{fdbken} 
		\end{align}
		which transmits the message $ m \in \{1, 2, \dots ,|\mathscr{I}|\} $ across the channel. The zero-error feedback code constructed above is uniform as well. 
		
		{By repeating the above argument for any starting time $ t_0+1>1 $, \eqref{fdbken} yields a uniform zero-error feedback code with the following change of variables. 
			\begin{align}
				\begin{split}
					X^{new}_t &= X(t_0+t)-\hat{X}_{t_0},\\
					\hat{X}^{new}_t &= \hat{X}_{t_0+t}-\hat{X}_{t_0}\\
					E^{new}_t &= E_{t_0+t},\\
					Q^{new}_t&= Q_{t_0+t},\\
					Y^{new}_t &= Y_{t_0+t},\\
					V^{new}_t &= V_{t_0+t}.
				\end{split} \label{chvar}
			\end{align}
		Note that, by \eqref{zef}, the coding function does not depend on $ t_0 $.
		Here, $ \hat{X}_{t_0} $ is the state estimate at the end of previous communication, and by above construction, $ X^{new}_t $ belongs to a ball with a known non-zero radius.}
		
		Therefore, bounded estimation error guarantees the existence of a uniform zero-error feedback code with a rate $ R $ that  satisfies
		\begin{align}
			R &= \frac{1}{\nu-1}\log |\mathscr{I}|\\
			&>  \kappa \log(1-\epsilon) -\frac{\kappa}{\nu-1}+\sum_{i=1}^{\kappa}\log|\lambda_i| .\label{istar}
		\end{align}
		By letting $ \nu \rightarrow \infty $ and the fact that $ \epsilon $ can be made arbitrarily small, the rate can be made close to $ h_{lin} $.	In other words, even for $ V_t= 0, \, \forall t \in \mathbb{N} $, to keep the plant states uniformly  bounded $R \geq h_{lin}$.
		
		By invoking the definition of the uniform zero-error feedback capacity (Definition \ref{def:c0f}), we get $ C_{0f} \geq R \geq h_{lin} $, which contradicts the assumption of the lemma. This demonstrates that in fact \eqref{estnec} holds.
		
		{Now, we drop the assumption $ V_t= 0, \, \forall t \in \mathbb{N} $ that we made earlier. So far, we have shown that if a coder-estimator achieves $ \sup \big\llbracket \norm{\bar{E}(t)}| V_{1:t}=0 \big\rrbracket \leq \infty$ then the uniform zero-error capacity of the channel has to satisfy $ C_0 \geq h_{lin} $. This can be rephrased as if $  C_0 < h_{lin} $ then for any coder-estimator,
			\begin{align}
				\sup \big\llbracket \norm{E_t} | V_{1:t}=0 \big\rrbracket= \infty. \label{infest}
			\end{align}
			In addition, as conditioning reduces the range \cite{nair2013nonstochastic}, we have 
			\begin{align*}
				\sup \big\llbracket \norm{E_t}\big\rrbracket &\geq \sup \big\llbracket \norm{E_t}| V_{1:t}=0 \big\rrbracket.
			\end{align*}
			Therefore, if \eqref{infest} holds then $ \sup \big\llbracket \norm{E_t}\big\rrbracket= \infty$.}
	\end{proof}
	In other words, if $  C_0 < h_{lin} $ then the worst-case estimation error is not uniformly bounded. By this, the proof of necessity is complete.
	
	\subsubsection{Achievability}
	Pick numbers $ \varsigma $, $ R $, and $ \gamma $ such that
	\begin{align}
		C_{0f} > R > \varsigma > h_{lin}\quad \text{and} \quad \gamma > \norm{A}.
	\end{align}
	By employing Lemma \ref{lem:lvl}, we pick a large enough $ r $, an $ r- $contracted quantizer $ \mathscr{Q}_r $ in $ \mathbb{R}^{n_x} $ with the contraction rate $ \rho_{\mathscr{Q}} $ and $ 2^{rh_{lin}} < M \lesssim 2^{r\varsigma} $ levels\footnote{Note that by Lemma \ref{lem:lvl} for large enough $ r $, this is guaranteed.}, where
	$ \rho_{\mathscr{Q}} \in (0,1) $ and does not depend on $ r $. Moreover, by Lemma \ref{lem:delc0f} a uniform zero-error feedback code with rate $ R $ exists.
	
	The operation of the encoder and decoder is organized into epochs $ \left[ \tau_i:=ir+1, \tau_{i+1} \right), i=1,2,\dots $. In each epoch, a block code of length $ r $ can be used to transmit the quantizer outputs $ \bar{q} = \mathscr{Q}_r(x) $ without errors. {In other words, the encoder maps $ \bar{q} $ to the channel input alphabets of length $ r $ with a $ (2^{rR},r) $ uniform zero-error feedback code.} 
	
	The decoder computes a state estimate $ \hat{x}_t $ as well as an upper bound $ \delta_t $ of the estimation error at time $ t $. 
	These operations are duplicated at the encoder as well.
	
	The encoder employs the quantizer  $  \mathscr{Q}_r$ and computes the quantized value $ \bar{q}_{\tau_i} $ of the current {\em scaled estimation error},  $ \varepsilon_{\tau_i} $ at time $ \tau_i $ produced by the encoder–decoder pair:
	\begin{align}
		\bar{q}_{\tau_i}=\mathscr{Q}_r(\varepsilon_{\tau_i}),\quad\varepsilon_{\tau_i} := \frac{x_{\tau_i}-\hat{x}_{\tau_i}}{\delta_{\tau_i}}. \label{estqr}
	\end{align}
	And encodes it by means of the feedback encoding function with block length $ r $ and sends it across the channel during the next epoch $ \left[ \tau_i, \tau_{i+1} \right)  $. 
	
	At the decoder, the error-less decoding rule is applied to the data received within the previous epoch $ \left[ \tau_{i-1}, \tau_{i} \right) $ and therefore computes the quantized and scaled estimation error $ \bar{q}_{\tau_{i-1}} $.
	Next, the estimate and the exactness bound is updated:
	\begin{align}
		\hat{x}_{\tau_i} &= A^r \, \big( \hat{x}_{\tau_{i-1}} + \delta_{\tau_{i-1}} \bar{q}_{\tau_{i-1}}\big),\label{estimup}\\
		\delta_{\tau_i} &= \delta_{\tau_{i-1}} \rho_{\mathscr{Q} }+\delta_*,\label{deltaup}
	\end{align}
	where $ \delta_*>0 $ is a constant scalar and $ \rho_{\mathscr{Q}}  $ is the contraction rate of the quantizer $ \mathscr{Q}_r $.
	
	The encoder and decoder are given common and arbitrarily chosen values $ \hat{x}_1=0 $, $ \delta_{\tau_1}> D_x$, and $ \delta_* > D \gamma^r $. 
	
	\begin{lem} \label{lem:est}
		The coding method introduced above keeps the estimation error uniformly bounded.
	\end{lem}
	\begin{proof}
		Let $ \delta_i:= \delta_{\tau_i} $, using \eqref{deltaup} and $ \rho_{\mathscr{Q}} \in (0,1)$ we have 
		\begin{align}
			\delta_i &= \rho_{\mathscr{Q}}^{i}\delta_1+\delta_*\frac{1-\rho_{\mathscr{Q}}^{i}}{1-\rho_{\mathscr{Q}}} < \delta_1+ \frac{\delta_*}{1-\rho_{\mathscr{Q}}} ,\quad \forall i\in \mathbb{N}\, . \label{deltb}
		\end{align}
		Let $ \omega_i:=\norm{ \hat{x}_{\tau_i}-x_{\tau_i}}  $, and
		\begin{align}
			D_r:= D \norm{\sum_{\theta=1}^{r-1}A^{\theta}}. \label{dr}
		\end{align} 
		We show $ \{\omega_i\}_{i\geq 1} $ is uniformly bounded.
		\begin{align}
			\omega_i &= \norm{A^r ( \hat{x}_{\tau_{i-1}} + \delta_{\tau_{i-1}} \bar{q}_{\tau_{i-1}} -x_{\tau_{i-1}}) -\sum_{\theta=\tau_{i-1}}^{\tau_i-1}A^{\tau_i-1-\theta}v_{\theta}} \nonumber \\
			&\leq \norm{ A^r \delta_{\tau_{i-1}} \bigg(  \bar{q}_{\tau_{i-1}} - \frac{x_{\tau_{i-1}}-\hat{x}_{\tau_{i-1}}}{\delta_{\tau_{i-1}}}\bigg) } +\norm{\sum_{\theta=1}^{r-1}A^{r-1-\theta}v_{\theta}} \nonumber\\
			&\leq \delta_{\tau_{i-1}}\norm{ A^r  (\varepsilon_{\tau_{i-1}} - \bar{q}_{\tau_{i-1}}) }+ D_r \nonumber\\
			&\stackrel{\text{Def.} \ref{def:rcont}}{\leq} \delta_{i-1} \rho_{\mathscr{Q}}  + D_r. \label{upz1}
		\end{align}
		Now, we show that the assumption in Definition \ref{def:rcont}, i.e., $ \varepsilon_{\tau_i} \in \mathbf{B}_1, \, \forall i\in \mathbb{N} ,$ is valid. In other words,	$ |\varepsilon_{\tau_i}| < \delta_i $. 
		We show this by induction. Given $ \hat{x}_1=0 $, we have $ \omega_1=\norm{x_1}<D_x<\delta_1 $, hence $ \norm{\varepsilon_{\tau_1}}<1 $. Therefore, \eqref{estqr} can be used and, by Definition \ref{def:rcont}, $ \norm{ A^r  (\varepsilon_{\tau_{1}} - \bar{q}_{\tau_{1}}) } < \rho_{\mathscr{Q}} $. Now, assuming $ \norm{\varepsilon_{\tau_{i-1}}}<1 $, we have $ \norm{ A^r  (\varepsilon_{\tau_{i-1}} - \bar{q}_{\tau_{i-1}}) } < \rho_{\mathscr{Q}} $. Therefore, from \eqref{deltaup} and \eqref{upz1} we have
		\begin{align*}
			\norm{\varepsilon_{\tau_i}}= \frac{\omega_i}{\delta_i} & \leq \frac{\delta_{i-1} \rho_{\mathscr{Q}}  + D_r}{\delta_{i-1} \rho_{\mathscr{Q}}  + \delta_*} \\
			&< \frac{\delta_{i-1} \rho_{\mathscr{Q}}  + D \gamma^r}{\delta_{i-1} \rho_{\mathscr{Q}}  + D \gamma^r} = 1.
		\end{align*}
		In other words, by choosing $ \delta_* $ large enough and according to the update rules, $ \delta_i $ is an upper bound for the estimation error. 	From \eqref{deltb} and \eqref{upz1}, we have
		\begin{align}
			\omega_i &< \rho_{\mathscr{Q}} \bigg( \delta_1+ \frac{\delta_*}{1-\rho_{\mathscr{Q}}} \bigg)  + D_r. \label{upz}
		\end{align}
		Therefore, the estimation error is uniformly bounded.
	\end{proof}
	\section{Proof of Theorem \ref{thm:stab} (stabilization condition)} \label{app:stab}
	The proof of the converse and achievability are given separately in the following.
	\subsubsection{ Converse}
	We adopt the approach in \cite{tatikonda2004controlb}, where the stabilization problem is converted to an estimation problem. {Assume that there exists a coder-controller such that the system \eqref{lti1} is uniformly bounded. We assume that the system is initialized at time $ 1 $, i.e., $ t_0=0 $. However, similar to the estimation problem, the results hold for any starting time $ t_0+1 \in \mathbb{N} $.
		For a given control sequence $ U_{1:t-1} $,} we have	
	\begin{align*}
		X_{t}=A^{t-1}X_1-\alpha_t(U_{1:t-1})+\beta_t(V_{1:t-1}),
	\end{align*}
	where 
	\begin{align*}
		\alpha_t(U_{1:t-1})&:=-\sum_{i=1}^{t-1}A^{t-1-i}BU_i,\\
		\beta_t(V_{1:t-1})&:=\sum_{i=1}^{t-1}A^{t-1-i}V_i.
	\end{align*}
	Bounded {stabilization} implies that 
	\begin{align*}
		\sup_{t \in \mathbb{N}}\|X_t\|&= \sup_{t \in \mathbb{N}}\norm{A^{t-1}X_1-\alpha_t(U_{1:t-1})+\beta_t(V_{1:t})}\\
		&< \phi<\infty.
	\end{align*}
	{Here, $ \alpha_t $ can be seen as a reconstruction of {the uncontrolled system $ X^{un}_t := A^{t-1}{X_1}+\beta_t(V_{1:t-1}) $ with distortion $ <\phi $. Define $ \hat{X}^{un}_t := \alpha_t(U_{1:t-1}) $.
	Therefore, the estimation error $\sup_{t \in \mathbb{N}}\|\hat{X}^{un}_t  - X^{un}_t\| = \sup_{t \in \mathbb{N}}\|X_t\| <\phi$.}
		By Theorem \ref{thm:finitem}, a necessary condition to achieve this condition is that the uniform zero-error feedback capacity of the connecting channel is not smaller  than the LTI system's topological entropy.}
	
	{We show that the uniform zero-error capacity of this connecting channel is upper bounded by $ C_{0f} $ of the original channel (i.e., the mapping from $ \{Q_t\} $ to $ \{Y_t\} $).  Note that there is feedback from the channel output, and the coder-controller can utilize it to construct a code.} Figure \ref{fig:eqset} shows an equivalent structure for the control problem (see Fig. \ref{fig:ctrl}) based on the above discussion. The dashed red box can be considered as the new encoder. The capacity of the resultant channel is upper bounded by the zero-error capacity of the original channel with noiseless feedback as using this potentially non-optimal encoder can only achieve the rate below feedback capacity. Therefore using Theorem \ref{thm:finitem} we have
	\begin{align*}
		C_{0f} \geq h_{lin}.
	\end{align*}
	\tikzstyle{sum} = [draw, fill=white, circle]
	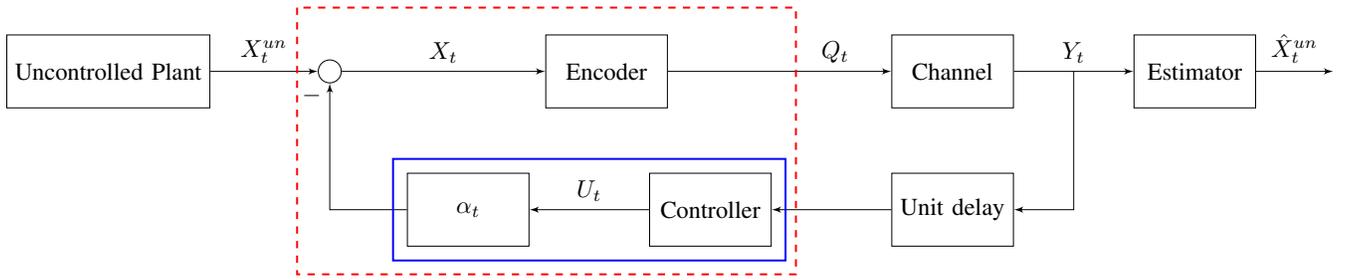
\begin{figure*}[t]
		\centering
		\scalebox{0.92}{\begin{tikzpicture}[auto, node distance=2cm,>=latex']
				\node [block2] (plant) { Uncontrolled Plant };
				\node [sum, right of=plant, node distance=3.2cm] (sum) {};
				\node [block1, right of=sum, node distance=4cm] (encoder) { Encoder };
				\node [block2, right of=encoder, node distance=5cm] (channel) { Channel };
				\node [block1, right of=channel, node distance=3.5cm] (decoder) { Estimator} ;
				\node [output, right of=decoder] (output) {};
				\node [block2, below of=channel] (controlsignal) {Unit delay };
				\node [block1, left of=controlsignal, node distance=3.5cm] (dec2) {Controller};
				\node [block2, left of=dec2, node distance=3.5cm] (alpha) {$ \alpha_t$};
				
				\draw [draw,->] (plant) -- node {$ X^{un}_t $} (sum);
				\draw [->] (sum) -- node {$ X_t $} (encoder);
				\draw [->] (encoder) -- node [pos=.75] {$ Q_t $} (channel);
				\draw [->] (channel) -- node [name=r] {$ Y_t $} (decoder);
				\draw [->] (decoder) -- node [name=u] {$ \hat{X}^{un}_t $}(output);
				\draw [->] (r) |- (controlsignal);
				\draw [->] (controlsignal) -- node {} (dec2);
				\draw [->] (dec2) -- node[above] {$ U_t $} (alpha);
				\draw [->] (alpha) -| node[pos=0.95] {$-$} (sum);
				\draw[thick,red,dashed] ($(sum.north west)+(-0.35,0.8)$)  rectangle ($(dec2.south east)+(0.35,-0.4)$);
				\draw[thick,blue] ($(dec2.north east)+(0.2,0.2)$)  rectangle ($(alpha.south west)+(-0.2,-0.2)$);	
		\end{tikzpicture}}
		\caption{Reduction of the stabilization problem to an estimation problem with channel feedback. The dashed red box is the feedback encoder of the new structure of the channel. The solid blue box is a copy of the estimator.}
		\label{fig:eqset}
	\end{figure*}
	\subsubsection{ Achievability}
	We first give the following lemma.
	\begin{lem}\label{lem:signaling}
		For the LTI system \eqref{lti1} and any $ N=1,2,... $, there exists a set of $ N $ controls
		\begin{align*}
			\mathcal{U}=\big\{u(j)| u(j) \in \mathbb{R}^{n_u}, j=1,\dots,N\big\} ,
		\end{align*}
		such that by considering $ U_{t} = \bar{u}+u(\nu) $, where $ \bar{u} $ is known {at the encoder} but $ \nu $ is not known, the unknown $ \nu $ can be determined without error for given $ X_{t+1},\, X_t $, $ \bar{u}  $, and having
		\[ \norm{ Bu(\nu) -Bu(\nu')} > 2D, \forall \nu \neq \nu' \in \{1,\dots,N\}.\]
	\end{lem}
	\begin{proof}
		Consider a state estimate at the encoder is given by 
		\begin{align}
			\tilde{X}_{t+1}:=A X_{t}+B\bar{u}. \label{uupd}
		\end{align}
		Therefore, by considering first $ U_t=\bar{u} $ (for now $ u(\nu)=0 $) and applying it in \eqref{lti1}, we obtain
		\begin{align}
			\norm{ X_{t+1}-\tilde{X}_{t+1}} &= \norm{ A X_{t}+B\bar{u}+V_{t}-A X_{t}-B\bar{u}} \nonumber \\
			& \leq D , \label{ballv}
		\end{align}
		where $ \norm{V_t}\leq D $ by assumption A2. Next, assume $ U_{t}=\bar{u}+u(\nu) $ where $ u(\nu) $ is not known at the encoder and updates $ \tilde{X}_{t+1} $ according to \eqref{uupd}. Hence, by the same argument above (with $ u(\nu) \neq 0 $)
		\begin{align*}
			\norm{ X_{t+1}-\tilde{X}_{t+1}-Bu(\nu)} \leq D .
		\end{align*}
		Thus $ \norm{ X_{t+1}-\tilde{X}_{t+1}} $ builds a ball of radius $ D $ with center $ \vartheta(\nu) := Bu(\nu) $.
		Let pick the set $ \{u(\nu)\} $ so that points in $ \{\vartheta(\nu)\} $ be $ 2D $ separated, i.e., $ \norm{ \vartheta(\nu) -\vartheta(\nu')} > 2D$ if $\nu \neq \nu' $. Therefore, an error-less communication can be executed via the plant input to the plant output.
	\end{proof}
	Pick numbers $ \varsigma $ and $ R $ such that
	\begin{align*}
		C_{0f} > R > \varsigma > h_{lin}.
	\end{align*}
	By employing Lemma \ref{lem:lvl}, we pick large enough $ r $, an $(r-n_x)-$contracted quantizer $ \mathscr{Q}_{r-n_x} $ in $ R^{n_x} $ with the contraction rate $ \rho_{\mathscr{Q}} $ and $2^{(r-n_x) h_{lin}}< M \lesssim 2^{(r-n_x)\varsigma} $ levels. Due to controllable $ (A,B) $, there exists a a linear transformation $\mathcal{T}: \mathbb{R}^{n_x} \to \mathbb{R}^{n_u \times n_x}$, called {\em deadbeat stabilizer} that can take state from any initial point $ x_1 \in \mathbb{R}^{n_x} $ to $ x_{n_x}=0 $ in $ n_x $ time steps assuming $ V_t = 0, \, \forall t \in \mathbb{N}$ \cite[Ch. 3]{matveev2009estimation}.
	
	The operation of the encoder and decoder is organized into epochs $ \left[ \tau_i:=ir+1, \tau_{i+1} \right), i=1,2,\dots $ . In each epoch, a block code of length $ r-n_x $ can be used to transmit the quantizer outputs $ q = \mathscr{Q}_{r-n_x}(x) $ in $ r-n_x $ transmissions without error. Then $ n_x $ time instants in the epoch are not used for transmissions. This will be made clear later.
	
	This transmission needs a feedback communication of $ y \in \mathcal{Y} $ from the decoder to the encoder. To arrange for this, we employ Lemma \ref{lem:signaling} and pick a feedback control alphabet of size $ \big|\mathcal{Y}\big| $. Its elements are labeled by the channel output letters $ \{u(y)\}_{y \in \mathcal{Y}} $.
	
	The encoder and decoder compute control command $  u_t $ which is produced as the sum $ u_t = u^b_t + u^{c}_t $ of two parts. The basic control $ u^b_t $ aims to stabilize the plant, whereas the communication control $ u^{c}_t $ serves the feedback communication of $ y_t $ from the decoder to the encoder. The basic controls are
	generated at times $ \tau_i $ in the form of a control program for the entire operation cycle $ [\tau_i : \tau_{i+1}) $.
	The current communication control is generated at the current time $ t $ on the basis of the message $ y_t $ currently received over the channel: $ u^{c}_t = u(y)$. This ensures unit delayed communication of $ y_t $ to the encoder, as is required by the block code at hand. The encoder employs this code to transmit the quantized value of the {\em scaled state} (see \eqref{scs}). This value is determined at the beginning of the operation cycle $ \tau_i $ and transmit if during the cycle $ [\tau_i : \tau_{i+1}) $. However since the length $ r-n_x  $ of the block
	code is less than the cycle duration $ r $, the transmission will be completed at time $ \tau_{i+1} - n_x - 1 $. Let $ \tau^*_{i}:=\tau_{i+1}-n_x $, during the remainder 
	$ [\tau^*_{i} : \tau_{i+1}) $, the encoder sends nothing over the channel. Hence for $ t \in [\tau^*_{i} : \tau_{i+1}) $, there is no need to communicate $ y_t $ from the decoder to the encoder and thus no need to employ communication control. The decoder uses this time to cancel the influence of the previously generated sequence of communication controls $ u^{c}_{\tau_i}, \dots , u^{c}_{\tau_{i+i}-n_x-1}$
	on the plant. To this end, for $ t \in [\tau^*_{i} : \tau_{i+1})  $ it puts 
	\begin{align}
		u^{c}_{\tau^*_{i}:\tau_{i+1}-1}=\mathcal{T}\big( \Xi(u_{\tau_{i}:\tau^*_{i}-1}) \big), \label{ct}
	\end{align}
	where $ \Xi(u_{\tau_{i+1}:\tau^*_{i}-1}) \in \mathbb{R}^{n_x} $ is the accumulated influence of the previous communication controls and can be obtained by 
	\begin{align*}
		\Xi(u_{\tau_{i}:\tau^*_{i}-1}) & :=\sum_{\theta=\tau_i}^{\tau^*_{i}-1} A^{\tau^*_{i}-1-\theta} B u^{c}_{\theta}.
	\end{align*}
	The encoder generates the control commands $ u^{c}_t $ so that they be replicas of $ u_t $. To this end, it calculates the basic controls by itself with overtaking the decoder by one cycle. For $ t \in [\tau_{i} : \tau^*_{i})  $, it gets aware of $ y_t $ and thus $ u^{c}_t $
	at time $ t + 1 $. So at time $ \tau^*_{i} $, the encoder is able to determine the ``canceling tail” \eqref{ct}. 
	
	The encoder takes the following actions:
	\begin{itemize}
		\item Computes the  quantized value of the {\em scaled state}:  
		\begin{align}
			\bar{q}_{\tau_i}=\mathscr{Q}_{r-n_x}(\varepsilon_{\tau_i}),\quad\varepsilon_{\tau_i}:= \frac{x_{\tau_i}}{\delta_{\tau_i}}. \label{scs}
		\end{align}
		And encodes the quantized scaled state by the feedback encoding function with block length $ r $ and sends it across the channel during the next epoch $ \left[ \tau_i, \tau_{i+1} \right)  $.
		\item Computes the basic control over the next epoch $ \left[ \tau_{i+1}, \tau_{i+2} \right)  $ and updates the state upper bound:		
		\begin{align*}
			u^b_{\tau_{i+1}:\tau_{i+2}-1} &=  \,  \mathcal{T}(\hat{x}_{\tau_{i+1}|\tau_i}) ,\\
			\delta_{\tau_i} &= \delta_{\tau_{i-1}} \rho_{\mathscr{Q}}+\delta_*,
		\end{align*}
		where $ \delta_1>\norm{\llbracket X_1 \rrbracket}$, $ \delta_* > D \gamma^r  $ are arbitrarily chosen constants and $ \rho_{\mathscr{Q}}  $ is the contraction rate of the quantizer $ \mathscr{Q}_{r-n_x} $. Here, 
		\begin{align}
			\hat{x}_{\tau_{i+1}|\tau_i}=\delta_{\tau_i}A^r \bar{q}_{\tau_{i}}+\sum_{\theta=\tau_{i}}^{\tau_{i+1}-1}A^{\tau_{i+1}-1-\theta} B u^b_{\theta}, \label{stest}
		\end{align}
		which is the estimated value of states at $ \tau_{i+1} $ based on the current measurements. For the first epoch, we set $ u^b_{1:r}=0 $.
	\end{itemize}	
	The decoder takes the following actions:
	\begin{itemize}
		\item Decodes the received block code and hence obtains  $ q_{ \tau_{i-1}} $ at time $ \tau_i $.
		\item Computes the basic control for the epoch $ \left[ \tau_{i}, \tau_{i+1} \right)  $ and updates the state upper bound:
		\begin{align*}
			u^b_{\tau_{i}:\tau_{i+1}-1} &=  \mathcal{T}(\hat{x}_{\tau_{i}|\tau_{i-1}}) ,\\
			\delta_{\tau_{i-1}} &= \delta_{\tau_{i-2}} \rho_{\mathscr{Q}}+\delta_*,
		\end{align*}
		where, $ \hat{x}_{\tau_{i}|\tau_{i-1}} $ is obtained similar to \eqref{stest} with one epoch delay due to the communication. 
		Note that because of the error-less transmission, $ \bar{q}_{\tau_{i-1}} $ is available at the decoder (with one epoch delay) and so $ u^b_{\tau_{i}:\tau_{i+1}-1} $, $ \delta_{\tau_{i-1}} $, as well as $ \hat{x}_{\tau_{i}|\tau_{i-1}} $ are reconstructed with the same values at the encoder.
		\item Calculates the communication control:		
		\begin{align*}
			u^{c}_t&
			=\begin{cases}
				u_t(y) \text{, }\qquad\qquad\qquad\quad\quad\quad\quad\quad \tau_i \leq t < \tau^*_{i} \\
				(t-\tau^*_{i}+1)\text{-th column of \eqref{ct}, }\,\,\, \tau^*_{i} \leq t < \tau_{i+1}
			\end{cases}
		\end{align*}
		\item Applies $ u_t = u^b_t + u^{c}_t  $.
	\end{itemize}
	\begin{figure*}[t]
		\centering
		\includegraphics[width=130mm]{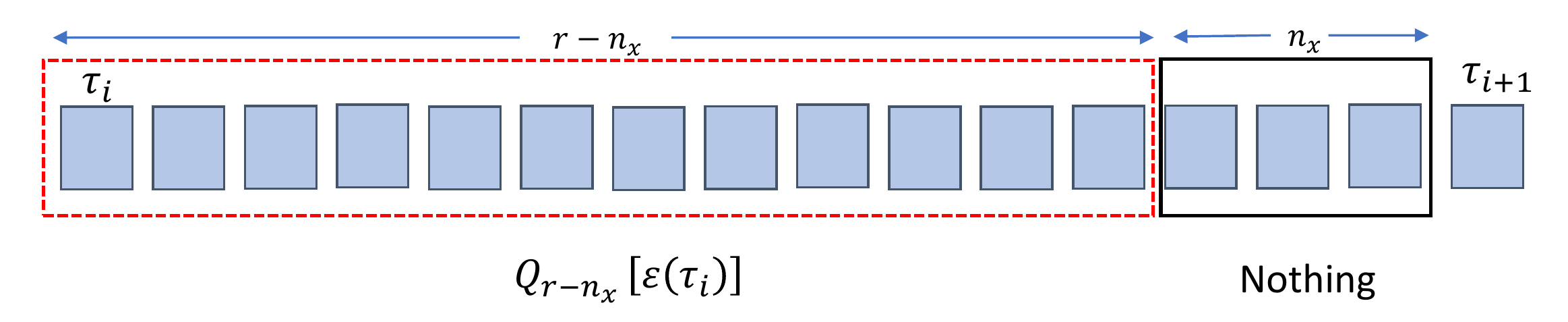}
		\caption{Time-line structure for the channel input.}
		\label{fig:tl2}
	\end{figure*}
	\begin{lem} \label{lem:stab}
		The coding method introduced above keeps the states bounded.
	\end{lem}
	\begin{proof}
		We first show that the state estimate in \eqref{stest} has a bounded error. We have	
		\begin{align}
			\begin{split}
				x_{\tau_{i-1}}-\hat{x}_{\tau_{i-1}|\tau_{i-2}} = &A^r \big(x_{\tau_{i-1}}-\delta_{\tau_i} \bar{q}_{\tau_{i}} \big) \\
				&+\sum_{\theta=\tau_{i-2}}^{\tau_{i-1}-1}A^{\tau_{i-1}-1-\theta}\big(B u_{\theta}-B u^b_{\theta}+v_{\theta}\big).
			\end{split}\label{esterr}
		\end{align}
		Note that $ u_{\theta}- u^b_{\theta}=u^c_{\theta} $ and because the communication control impact is canceled in the last $ n_x $ steps of the epoch, we have				
		\begin{align}
			\sum_{\theta=\tau_{i-2}}^{\tau_{i-1}-1}A^{\tau_{i-1}-1-\theta} B u^{c}_{\theta}=0. \label{comcon}
		\end{align}	
		Let $ \delta_{i-1}:=\delta_{\tau_{i-1}} $ and $ D_r $ be the upper bound on the {noise} effect, defined in \eqref{dr}. By considering \eqref{comcon} and substituting  \eqref{scs} in \eqref{esterr}, we obtain		
		\begin{align}
			\norm{x_{\tau_{i-1}}-\hat{x}_{\tau_{i-1}|\tau_{i-2}}} &\leq  \delta_{i-1} \norm{ A^r (\varepsilon_{\tau_{i-1}}-\bar{q}_{\tau_{i-1}}) } +D_r. \label{estnorm}
		\end{align}
		Note that since the updating rule for $ \delta_{i}$ is the same as \eqref{deltaup}, hence \eqref{estnorm} admits the same bound in \eqref{deltb} and therefore
		\begin{align*}
			\norm{x_{\tau_{i-1}}-\hat{x}_{\tau_{i-1}|\tau_{i-2}}} &\leq  \rho_{\mathscr{Q}} \bigg( \delta_1+ \frac{\delta_*}{1-\rho_{\mathscr{Q}}} \bigg)  + D_r.
		\end{align*}
		We now consider the state evolution and then 
		\begin{align}
			\norm{x_{\tau_i}} &= \norm{ A^r x_{\tau_{i-1}} +\sum_{\theta=\tau_{i-1}}^{\tau_{i}-1}A^{\tau_{i}-1-\theta}\big(B u _{\theta}+v_{\theta}\big)} \nonumber \\
			&\leq \norm{ A^r x_{\tau_{i-1}} +\sum_{\theta=\tau_{i-1}}^{\tau_{i}-1}A^{\tau_{i}-1-\theta} B \big(u^b _{\theta}+u^{c}_{\theta}\big) } +D_r \nonumber\\
			\begin{split}
				&\leq \norm{ A^r \big(x_{\tau_{i-1}}-\hat{x}_{\tau_{i-1}|\tau_{i-2}}\big)}+D_r  \\
				&\qquad+ \norm{ A^r \hat{x}_{\tau_{i-1}|\tau_{i-2}} +\sum_{\theta=\tau_{i-1}}^{\tau_{i}-1}A^{\tau_{i}-1-\theta} B \big(u^b _{\theta}+u^{c}_{\theta}\big) } .
			\end{split} \label{upb1}
		\end{align}
		Since the deadbeat stabilizer, i.e., $ \mathcal{T}(\hat{x}_{\tau_{i-1}|\tau_{i-2}}) $ takes the states from $ \hat{x}_{\tau_{i-1}|\tau_{i-2}} $ to zero and considering that $ u^b_{\tau_{i-1}:\tau_{i}-1} = \mathcal{T}(\hat{x}_{\tau_{i-1}|\tau_{i-2}}) $, we have
		\begin{align}
			A^r \hat{x}_{\tau_{i-1}|\tau_{i-2}} + \sum_{\theta=\tau_{i-1}}^{\tau_{i}-1} A^{\tau_{i}-1-\theta} B u^b_{\theta}=0. \label{bascon}
		\end{align}	
		Here, similar to \eqref{comcon}, $ \sum_{\theta=\tau_{i-1}}^{\tau_{i}-1}A^{\tau_{i}-1-\theta}B u^c_{\theta}=0 $. Thereby,
		\begin{align*}
			\norm{x_{\tau_i}} &\leq A^r\norm{x_{\tau_{i-1}}-\hat{x}_{\tau_{i-1}|\tau_{i-2}}}+D_r\\
			&\leq \norm{A^r} \rho_{\mathscr{Q}} \bigg( \delta_1+ \frac{\delta_*}{1-\rho_{\mathscr{Q}}} \bigg)  + D_r(1+\norm{A^r}),
		\end{align*}
		which shows that the plant state is uniformly bounded and this completes the proof.
	\end{proof}	
	\section{Proof of Theorem \ref{thm:c0f0}~(condition on $C_0=0$)}\label{app:c0f0}
	\textit{Sufficiency:} We show that for any choice of encoding functions and blocklength $ n $ there is a common output for $ m,m' \in \mathcal{M}$, i.e., $ \exists z_{1:n},z'_{1:n} $ such that the output sequences, $ y_{1:n}=y'_{1:n} $, where $ y_{1:n}=f_{1:n}\big(m,z_{1:n-1}\big)\oplus z_{1:n}, \,y'_{1:n}=f_{1:n}\big(m',z'_{1:n-1}\big)\oplus z'_{1:n} $.\footnote{Here, with a slight abuse of notation, it is assumed that $ f_{i}\big(m,z_{1:i-1}\big)= f_{i}\big(m,y_{1:i-1}\big) $.} In other words, $\forall \, n \in \mathbb{N}$, and  
	\begin{align*}
		d_{1:n}:=f_{1:n}\big(m',z'_{1:n-1}\big) \ominus f_{1:n}\big(m,z_{1:n-1}\big)\, \in \mathcal{X}^n,
	\end{align*}
	$\exists\, z_{1:n},z'_{1:n}$ such that $d_{1:n}= z_{1:n} \ominus z'_{1:n} $.
	
	First observe that having current states $ S_i=s$ and $S'_i=s' $, for two noise sequences of $ z_{1:i-1}$ and $z'_{1:i-1} $, respectively, the label on out-going edges in the coupled graph is belong to $ \{z_i \ominus z'_i |  S_i=s,  S'_i=s' \}$. 
	Now consider the first transmission, by choosing any inputs $ f_{1}(m), f_{1}(m') \in \mathcal{X}$, if there is an edge from any state $ (k,j) \in V $ with the value $d_1:=f_{1}(m')\ominus f_{1}(m) \in \mathcal{X}$ then there exist $ z_1,z'_1 \in \mathcal{X} $ that produce a common output for two channel inputs $ f_{1}(m) $ and $ f_{1}(m') $. By continuing this argument for any $ i\in \mathbb{N} $ having $ y_{1:i-1}=y'_{1:i-1} $, if $ d_i =f_{i}(m',z'_{1:i-1}) \ominus f_{i}(m,z_{1:i-1}) \in \mathcal{X} $ is chosen such that there is an edge with value $ d_i $ then there is an output shared with two messages. In other words, by choosing any value for $ d_i $, if there is an edge with corresponding value it means there is a pair of noise values $ (z_i,z'_i) $ such that $ d_i=z_i \ominus z'_i $, therefore $ y_i=y'_i $. If there is no such an edge for a particular $ d_i $, then there is no pair of noise values that produces the same output, and thus, $  y_i \neq y'_i $.
	
	Therefore, if $ \forall n \in \mathbb{N} $ and for any choice of $ d_{1:n} \in \mathcal{X}^n $ there is a walk on the coupled graph then the corresponding noise sequences of the walk can produce the same output, i.e., $ y_{1:n}=y'_{1:n} $ which implies $ C_{0f}=0 $ and therefore $ C_0=0 $.\\
	
	\textit{Necessity:} Assume there is no walk for a sequence of $ d_{1:n} $ then by choosing any two input sequences $x_{1:n}, x'_{1:n}$ such that $ x_{1:n}\ominus x'_{1:n}=d_{1:n} $, two messages $ m $ and $ m' $ can be transmitted with zero-error which contradict with the assumption that $ C_{0}=0 $ (and also $ C_{0f}=0 $).
	
	\section{Proof of Lemma \ref{lem:out}} \label{app:out}
	The output sequence, $ y_{1:n} $, is a function of input sequence, $ x_{1:n} $, and channel noise, $ z_{1:n} $, which can be represented as the following
	\begin{align}
		y_{1:n} = x_{1:n} \oplus z_{1:n},  \label{adnoise}
	\end{align}
	where $ z_{1:n} \in \mathcal{Z}(s_1,n) $. The set of all output sequences $ \mathcal{Y}(s_1,x_{1:n}) $ can be obtained as $\mathcal{Y}(s_1,x_{1:n})=\bigl\{ x_{1:n} \oplus z_{1:n}| z_{1:n} \in \mathcal{Z}(s_1,n)\bigr\}$.
	Since for given $ x_{1:n} $, \eqref{adnoise} is bijective, we have the following
	\begin{align}
		\big|\mathcal{Y}(s_1,x_{1:n})\big|=\big| \mathcal{Z}(s_1,n)\big|. \label{yvsize}
	\end{align}
	For a given initial state $s_1\in\mathcal{S}$, define the binary indicator vector $ \zeta \in\{0,1\}^{|\mathcal{S}|} $ consisting of all zeros except for a 1 in the position corresponding to $s_1$; e.g., in Fig.\ref{fig:mealy}, if starting from state $ S=0 $, then {$\zeta =[1, 0]$}. Observe that since each output of the finite-state additive channel triggers a different state transition, each sequence of state transitions has a one-to-one correspondence to the output sequence, given the input sequence.
	
	The total number of state trajectories after $n$-step starting from state $s_i$ is equal to sum of $i$-th row of $\mathcal{A}^n$ \cite{lind1995introduction}. Hence, because of a one-to-one correspondence between state sequences and output sequences then $|\mathcal{Z}(s_1,n)|=\zeta ^\top \mathcal{A}^n \mathbbm{1}$. 
	
	Next, we show the upper and lower bounds in \eqref{outlam}. According to the Perron-Frobenius Theorem, for an irreducible $|\mathcal{S}|\times |\mathcal{S}|$ matrix $\mathcal{A} $ (or, equivalently, the adjacency matrix for a strongly connected graph), the entries of eigenvector $v \in \mathbb{R}^{|\mathcal{S}|}$ corresponding to $\lambda$ are strictly positive \cite[Thm. 8.8.1]{godsil2001strongly},\cite[Thm. 4.2.3]{lind1995introduction}. Therefore, multiplying $ \mathcal{A} $ by $ \mathcal{A} v=\lambda v $ results in $ \mathcal{A}^n v=\lambda^n v $ for $ n \in \mathbb{N} $.
	Left multiplication by the indicator vector, $ \zeta ^\top $ yields
	\begin{align}
		\zeta ^\top\mathcal{A}^n v = \lambda^n \zeta ^\top v. \label{eigv}
	\end{align}
	Denote minimum and maximum element of vector $ v $ by $ v_{min} $ and $ v_{max} $ respectively.	Hence, considering that all the elements in both sides of \eqref{eigv} are positive, we have
	\begin{align}
		v_{min}\zeta ^\top\mathcal{A}^n \mathbbm{1}\leq \zeta ^\top\mathcal{A}^n v &\leq v_{max} \lambda^n \zeta ^\top \mathbbm{1} \nonumber \\
		& = v_{max} \lambda^n,
	\end{align}
	where $ \mathbbm{1} $ is all-one column vector. Therefore, dividing by $v_{min}$, we have
	\begin{align}
		\big|\mathcal{Y}(s_1,x_{1:n})\big| =\zeta ^\top\mathcal{A}^n &\leq \frac{v_{max}}{v_{min}} \lambda^n = \beta \lambda^n, \label{out_up}
	\end{align}
	where $ \beta:= v_{max}/v_{min} > 0 $.	Moreover, for deriving the lower bound similar to above, we have
	\begin{align*}
		v_{min}\lambda^n \zeta ^\top \mathbbm{1}\leq \zeta ^\top\mathcal{A}^n v &\leq v_{max} \zeta ^\top\mathcal{A}^n \mathbbm{1} \\
		& =v_{max} \big|\mathcal{Y}(s_1,x_{1:n}) \big|.
	\end{align*}
	Let  $ \alpha:= v_{min}/v_{max}=1/\beta > 0 $, hence $\alpha \lambda^n \leq \big|\mathcal{Y}(s_1,x_{1:n}) \big|$ which combining it with \eqref{out_up} results in \eqref{outlam}.
	
	\section{Proof of $ C_{0f} $ converse in Theorem \ref{thm:zero-error}} \label{app:zero-error}
	We prove no coding method can do better than \eqref{c0f}. 
	
	Let $ m \in \mathcal{M} $ be the message to be sent and  $ y_{1:n} $ be the output sequence received such that 
	\begin{align*}
		y_i=f_{i}(m,y_{1:i-1})\oplus z_i, \, i=1,\dots, n,
	\end{align*}
	where $ z_{1:n} \in \mathcal{Z}(s_1,n) \in \mathcal{X}^{n}$ is the additive noise and $ f_{i} $ the encoding function. Therefore, the output is a function of encoding function and noise sequence, i.e., \mbox{$ y_{1:n}=\psi(f_{1:n}(m),z_{1:n}) $}.
	We denote all possible outputs 
	\begin{align*}
		\Psi (\mathcal{F},\mathcal{Z}(s_1,n))=\bigl\{y_{1:n}| m \in \mathcal{M}, z_{1:n} \in \mathcal{Z}(s_1,n) \bigr\},
	\end{align*}
	where $ \mathcal{F}$ is a zero-error feedback code (which by Proposition \ref{prop:fsc0f} is also uniform). 
	
	For having a zero-error code any two $ m, m' \in \mathcal{M}, m\neq m'$ and any two $ z_{1:n}, z_{1:n}' \in \mathcal{Z}(s_1,n)$ must result in $ \psi(f_{1:n}(m),z_{1:n})\neq \psi(f_{1:n}(m'),z_{1:n}') $. Note that when $ m = m' $, (even with feedback) at first position that $ z_{1:n} \neq z_{1:n}' $ will result in $ \psi(f_{1:n}(m),z_{1:n})\neq \psi(f_{1:n}(m'),z_{1:n}') $. Therefore, assuming the initial condition is known at both encoder and decoder,
	\begin{align*}
		\big|\Psi (\mathcal{F},\mathcal{Z}(s_1,n))\big| =M\big|\mathcal{Z}(s_1,n)\big| \leq q^n.
	\end{align*}
	Therefore, $ M $ is an upper bound on the number of messages that can be transmitted when initial condition is not available. We know that 
	$ \alpha \lambda^n \leq \big|\mathcal{Z}(s_1,n)\big| \leq \beta \lambda^n $. Therefore, by Lemma  \ref{lem:cc0f}, we have
	\begin{align*}
		C_{0f} &= \lim_{n \to \infty} \sup_{ \mathcal{F} \in  \mathscr{F}(n)} \frac{\log |\mathcal{M}|}{n}  \\
		&\leq \lim_{n \to \infty } \frac{1}{n} \log \frac{q^n}{\alpha \lambda^n}\\
		&= \log q-\log \lambda .
	\end{align*}
	This proves the converse in \eqref{c0f}.		
	
	\section{Proof of Lemma \ref{lem:zcz}}\label{app:zcz}
	Assume $ C_0=0\,\, \forall n \in \mathbb{N} $. Therefore, for any input sequences $ x_{1:n},x'_{1:n} \in \mathcal{X}^n $, there exits at least one output sequence in common, i.e., $ \exists z_{1:n},z'_{1:n} \in \mathcal{Z}(n) $ such that 
	\begin{align}
		x_{1:n} \oplus z_{1:n} =x'_{1:n} \oplus z'_{1:n}.\label{eqzc0}
	\end{align}
	Hence, the set $ \big\{z_{1:n} \ominus z'_{1:n} | z_{1:n},z'_{1:n} \in \mathcal{Z}(n)\big\} $ must span the whole input space $ \mathcal{X}^n $. Otherwise, there exists $ p_{1:n} \notin \mathcal{P}(n) $. Now, choose two inputs of $ x_{1:n}=0 \dots 0 $ (all-zero sequence) and $x'_{1:n} = p_{1:n} $, since $ x_{1:n} \oplus z_{1:n} \neq p_{1:n} \oplus z'_{1:n} $ therefore $ x_{1:n} $ and $ x'_{1:n} $ are distinguishable which contradicts with $ C_0=0 $.
	
	On the other hand, if $\forall n \in \mathbb{N}, \, \mathcal{P}(n)=\mathcal{X}^n $, then for any input sequences $ x_{1:n},x'_{1:n} \in \mathcal{X}^n $, there exists at least one pair of noise sequences $ z_{1:n},z'_{1:n} \in \mathcal{Z}(n) $ such that 
	\begin{align*}
		x_{1:n} \ominus x'_{1:n} =z_{1:n} \ominus z'_{1:n},
	\end{align*}
	and therefore, $ C_0=0 $.
	
	\section{Proof of Lemma \ref{lem:win}}\label{app:win}
	By Definition \ref{subg}, $\mathcal{V}_A \subseteq \mathcal{V}_B $, thus, $ v_0 \in \mathcal{V}_B $. Moreover, for any outgoing edge from $ v_0 $ on  $\mathscr{G}_A $ there is an edge on $\mathscr{G}_B $. This argument holds for the remaining vertices associated with walk $\varpi_A(n,v_0)$. Therefore, a walk $ \varpi_B(n,v_0) $ can be constructed with the same sequence of edges associated with $ \varpi_A(n,v_0) $.
	\section{Proof of Proposition \ref{prop:extended0f0}}\label{app:extended0f0}
	
	Let $ \mathcal{Z}_A(n) $ and $ \mathcal{Z}_B(n) $ be sets of all possible noise sequences starting from any initial condition with length $ n $ for channel A and B, respectively. Let  $ \mathcal{Y}^{P}(x_{1:n})=\{x_{1:n} \oplus z_{1:n}| \mathcal{Z}^P(n)\} $ such that $ P \in \{A,B\} $ be the possible output set when the channel input is $ x_{1:n} $. Since $\mathscr{G}_A$ is a subgraph of $\mathscr{G}_B$, hence, by Lemma \ref{lem:win} $ \forall n, \, \mathcal{Z}_A(n) \subseteq \mathcal{Z}_B(n) $ and therefore
	\begin{align*}
		\mathcal{Y}^{A}(x_{1:n}) \subseteq \mathcal{Y}^{B} (x_{1:n}), \forall n \in \mathbb{N}.
	\end{align*}
	Which yields the number of distinguishable messages which can be transmitted from channel $ A $ is no larger than channel $ B $. 		
	\section{Proof of Proposition \ref{prop:ex0f0}} \label{app:ex0f0}
	By Theorem \ref{thm:c0f0} to have a $ C_0>0 $ there has to be a $ d(1:n) $ that does not admit a walk with the same label sequence on the coupled graph of the finite-state machine shown in Fig. \ref{fig:cg}. Any state except $ S=(1,1) $ has both outgoing edges with labels 0 and 1. Therefore, the only possible final state in a walk that does not admit any further edge with label 1 (it has only one outgoing edge with label 0) has to be state $ S=(1,1) $. The only state that has an edge to state $ S=(1,1) $ is $ S=(0,0) $. However, there are two edges with the same label 0 from this state. In other words, any walk leading to state $ S=(0,0) $ having label 0 for the next step, either can end up in state $ (0,0) $ or $ (1,1) $. Therefore, it is impossible to construct a sequence that does not admit a walk on the graph in Fig. \ref{fig:cg}.  Hence, $ C_0=0 $ and therefore $ C_{0f}=0 $.
	\section{$ C_{0f} $ of Example \ref{memless} using Shannon's formula} \label{app:shan}
	Let $ P_X $ be the channel input distribution, we have
	\begin{align*}
		C_{0f}&=\max_{P_X} \log\bigg(\max_{y_i \in \mathcal{X}} \sum_{z_i \in \mathcal{Z}}  P_X(y_i \ominus z_i)\bigg)^{-1}.
	\end{align*}
	This {\em minmax} optimization is equivalent to the following \textit{linear programming} problem.
	\begin{align}
		\begin{split}
			\min_{P_X, w} \quad & w\\
			\text{subject to} \quad  & \sum_{z_i \in \mathcal{Z}} P_X(y_i \ominus z_i) \leq w, \forall y_i \in \mathcal{X}.
		\end{split} \label{minmax2}
	\end{align}
	Let $ p:=\big(P_X(x_i)\big)_{x_i \in \mathcal{X}} $ be the input distribution vector. We define the Lagrangian associated with \eqref{minmax2} by
	\begin{align}
		\begin{split}
			L(p,w,\xi,\mu,\Gamma)=w-&\xi(\mathbbm{1}^Tp -1)-\mu^T p \, +\\
			& \sum_{y_i \in \mathcal{X}}\gamma(y_i)\bigg(\sum_{z_i \in \mathcal{Z}}  P_X(y_i \ominus z_i) -w\bigg),
		\end{split} \label{minmax23}
	\end{align}
	where $ \xi \in \mathbbm{R} $, $ \mu  \in \mathbbm{R}^{|\mathcal{X}|} $, and $ \Gamma:=\big(\gamma(y_i)\big)_{y_i \in \mathcal{X}} \in \mathbbm{R}^{|\mathcal{X}|} $ are Lagrange multipliers and $ \mathbbm{1} $ is a vector of all ones. The {\em Karush-Kuhn-Tucker }(KKT) conditions for \eqref{minmax23} are as follows	
	\begin{gather*}
		\frac{\partial L}{\partial \xi} =\mathbbm{1}^Tp-1=0,\\
		\frac{\partial L}{\partial w} = 1-\mathbbm{1}^T \Gamma=0 ,\\
		\frac{\partial L}{\partial p} = -\xi \mathbbm{1}-\mu+\varLambda\Gamma=0, \\
		\mu^T p = 0,\\
		\bigg(\sum_{z_i \in \mathcal{Z}}  P_X(y_i \ominus z_i)-w\bigg) \gamma(y_i)=0 , y_i \in \mathcal{X},\\
		\sum_{z_i \in \mathcal{Z}}  P_X(y_i \ominus z_i)\leq w , y_i \in \mathcal{X}\\
		\mu \succeq 0,\Gamma \succeq 0,p \succeq 0.
	\end{gather*}
	Here, $ \succeq 0 $ means all elements are non-negative and $ \varLambda $ is the adjacency matrix of the channel, where the element $ \varLambda_{jk}=1 $ if $ k- $th input can get mapped to $ j- $th output, otherwise $ \varLambda_{jk}=0 $. It is reasonably easy to see that the uniform distribution for input set  $ \big($i.e., $ P_X(x_i)=|\mathcal{X}|^{-1}, i=1,...,|\mathcal{X}| \big)$ is a solution to KKT conditions, yielding the unique minimum value for the cost function \cite[Ch. 5]{boyd2004convex}. We obtain $ \Gamma=|\mathcal{X}|^{-1} \mathbbm{1}$, $ \mu=0$, and $ w = \xi = |\mathcal{X}|^{-1}|\mathcal{Z}| $. 			
	Hence, $C_{0f}=-\log w $ gives \eqref{mlc0f}.
	
	\bibliographystyle{IEEEtran}
	\bibliography{T_IT}
	
			\begin{IEEEbiography} []
		{Amir Saberi} (S'18--M'22)
		Amir received the B.Sc. from the University of Tabriz, Iran in 2011, the M.Sc. from the University of Tehran, Iran in 2014, and the PhD from The University of Melbourne all in Electrical Engineering. Prior to his PhD, he had a 3-year career in industry working on anomaly detection in networked systems.  He is currently a postdoctoral fellow with The Australian National University. He has been the recipient of the Melbourne Research Scholarship in 2017 and the Student Engagement Grant in 2020 from The University of Melbourne. The main focus of his research includes Networked Control Systems, Information Theory, and Real-time Optimisation. 
	\end{IEEEbiography}
	\begin{IEEEbiography}[ ]
		{Farhad Farokhi} (S'11--M'15--SM'20) received PhD from the KTH Royal Institute of Technology in 2014. He joined The University of Melbourne, where he is currently a Lecturer (equivalent to Assistant Professor in North America). From 2018--2020, he was also a Research Scientist at the CSIRO's Data61. He has been the recipient of the VESKI Victoria Fellowship from the Victoria State Government, Australia, and the McKenzie Fellowship, the 2015 Early Career Researcher Award, and MSE Excellence Award for Early Career Research from The University of Melbourne. He has been involved in multiple projects on data privacy and cyber-security funded by the Australian Research Council, the Defence Science and Technology Group, the Department of the Prime Minister and Cabinet, the Department of Environment and Energy, and the CSIRO. He is the associate editor for IET Smart Grid, Results in Control and Optimization, and Conference Editorial Board of IEEE Control System Society.
	\end{IEEEbiography}
	\begin{IEEEbiography}[ ]
		{Girish N. Nair} (FIEEE) was born in Malaysia and is a Professor with the Department of Electrical and Electronic Engineering at The University of Melbourne. From 2015 – 2019 he was an ARC Future Fellow. He has received several prizes, including the IEEE CSS Axelby Outstanding Paper Award in 2014 and a SIAM Outstanding Paper Prize in 2006.
	\end{IEEEbiography}
\end{document}